\documentclass[12pt,preprint]{aastex}

\def\cm{\,{\rm cm}}

\def\erg{\,{\rm erg}}

\def\dd{{\rm d}}

\def\be{\begin{equation}}
\def\ee{\end{equation}}
\def\beq{\begin{eqnarray}}
\def\eeq{\end{eqnarray}}
\def\sT{\sigma_{\rm T}}

\def\Rload{R_{\rm load}}
\def\Rgap{R_{\rm gap}}
\def\Racc{R_{\rm acc}}

\def\macc{m_{\rm acc}}
\def\nuacc{\nu_{\rm acc}}
\def\tnuacc{\tilde{\nu}_{\rm acc}}

\def\mload{m_{\rm load}}
\def\Rdec{R_{\rm dec}}
\def\mdec{m_{\rm dec}}
\def\Eej{E_{\rm ej}}
\def\Gej{\Gamma_{\rm ej}}

\def\Grel{\Gamma_{\rm rel}}
\def\tGrel{\tilde{\Gamma}_{\rm rel}}
\def\tG{\tilde{\Gamma}}
\def\tg{\tilde{\gamma}}
\def\tnu{\tilde{\nu}}
\def\tep{\tilde{\epsilon}}
\def\tb{\tilde{\beta}}
\def\tP{\tilde{P}}
\def\tn{\tilde{n}}

\def\tB{\tilde{B}}
\def\tU{\tilde{U}}
\def\tf{\tilde{f}}
\def\tR{\tilde{R}}
\def\tm{\tilde{m}}
\def\tK{\tilde{K}}
\def\trho{\tilde{\rho}}

\def\Fnuobs{F_\nu}
\def\tobs{t_{\rm obs}}
\def\tdec{t_{\rm dec}}
\def\nuobs{\nu_{\rm obs}}
\def\nufluid{\nu_{\rm fluid}}
\def\Lobs{L^{\rm obs}}
\def\dLmax{\delta L^{\rm max}}

\def\epe{{\epsilon_e}}
\def\epB{\epsilon_B}
\def\tepB{\tilde{\epsilon}_B}

\def\tBej{\tilde{B}_{\rm ej}}
\def\Bej{B_{\rm ej}}

\def\xiacc{\xi_{\rm acc}}
\def\Zacc{Z_{\rm acc}}

\def\xiload{\xi_{\rm load}}
\def\xigap{\xi_{\rm gap}}

\def\mload{m_{\rm load}}

\def\Urad{U_{\rm rad}}

\newbox\grsign \setbox\grsign=\hbox{$>$} \newdimen\grdimen \grdimen=\ht\grsign
\newbox\simlessbox \newbox\simgreatbox \newbox\simpropbox
\setbox\simgreatbox=\hbox{\raise.5ex\hbox{$>$}\llap
     {\lower.5ex\hbox{$\sim$}}}\ht1=\grdimen\dp1=0pt
\setbox\simlessbox=\hbox{\raise.5ex\hbox{$<$}\llap
     {\lower.5ex\hbox{$\sim$}}}\ht2=\grdimen\dp2=0pt
\setbox\simpropbox=\hbox{\raise.5ex\hbox{$\propto$}\llap
     {\lower.5ex\hbox{$\sim$}}}\ht2=\grdimen\dp2=0pt
\def\simgt{\mathrel{\copy\simgreatbox}}
\def\simlt{\mathrel{\copy\simlessbox}}

\begin{document}

\title{Afterglow emission from pair-loaded blast waves in gamma-ray bursts}

\author{Andrei M. Beloborodov\altaffilmark{1}} 
\affil{Physics Department and Columbia Astrophysics Laboratory, \\
Columbia University, 538 West 120th Street New York, NY 10027}

\altaffiltext{1}{Also at Astro-Space Center, Lebedev Physical 
Institute, Profsojuznaja 84/32, Moscow 117810, Russia} 

\begin{abstract}
The MeV radiation front of gamma-ray bursts creates copious $e^\pm$ pairs 
as it propagates through an ambient medium. The created pairs enrich the 
leptonic component of the medium by a large factor at distances 
$R<\Rload\sim 10^{16}$~cm from the burst center. The following blast wave 
sweeps up the pair-rich medium and then emits the observed afterglow 
radiation. We find that the afterglow has a ``memory'' of $e^\pm$ loading 
outside $\Rload$. The $e^\pm$ remain in the swept-up material and slowly 
cool down by emitting synchrotron radiation. They are likely to dominate 
the blast-wave emission in IR, optical, and UV bands during the first 
minutes of the observed afterglow. The expected $e^\pm$ radiation is 
described by a simple formula, which is derived analytically and checked 
by numerical integration of synchrotron emission over the blast material;
a suitable Lagrangian formalism is developed for such calculations. 
The main signature of $e^\pm$ radiation is its flat (``white'') spectrum 
in a broad range of frequencies from IR to UV and possibly soft X-rays. 
This radiation can be detected by {\it Swift} satellite, which would 
enable new observational tests for the explosion physics.
\end{abstract}



\section{Introduction}

Cosmological gamma-ray bursts (GRBs) are produced by powerful explosions
in distant galaxies. It is not yet clear how the explosion is triggered,
however its basic phenomenological picture has been established: an 
ultra-relativistic shell (``fireball'') is ejected by a compact central 
engine. The expanding shell emits the burst of $\gamma$-rays, then sweeps 
up an ambient medium, and decelerates, producing the observed afterglow 
radiation. The afterglow is explained as synchrotron emission of nonthermal 
electrons in the relativistic blast wave (see Piran 2004 for a recent 
review).

Thompson \& Madau (2000) pointed out that an external medium must 
be $e^\pm$-loaded and preaccelerated by the leading $\gamma$-ray front
(prompt GRB radiation), which should affect the ensuing shock wave.
The effect can be thought of as   
a result of $\gamma$-ray transfer through the optically thin medium
(Beloborodov 2002, hereafter B02) which involves a runaway of pair 
creation and bulk acceleration.
A tiny fraction of GRB radiation participates in this transfer,
however, it impacts dramatically the circumburst medium.
The transfer problem was solved in B02, and
the number of loaded $e^\pm$ and the Lorentz factor of the medium behind
the $\gamma$-ray front were calculated. The loaded pairs were found to
dominate the ambient medium at radii 
$R<\Rload\approx 1.6\times 10^{16}(E_\gamma/10^{53}{\rm erg})$~cm,
where $E_\gamma$ is the isotropic ($4\pi$) equivalent of the GRB energy.

The $e^\pm$ loading sets the stage for the immediately following 
shock wave driven by the GRB ejecta. The shock heats and accelerates
the particles of the medium, and a nonthermal $e^\pm$ population is expected 
to form behind the shock front, which produces synchrotron radiation. 
The goal of the present paper is to calculate emission from the 
pair-loaded postshock plasma in the expanding blast wave. 

This difficult problem was previously approached in a few works.
B02 evaluated the shock parameters at $R<\Rload$ and found that the GRB 
afterglow should start with a brief and bright optical signal, however 
did not calculate the expected light curve or spectrum of $e^\pm$ radiation.
Then Li, Dai, \& Lu (2003) calculated the light curve considering the 
blast wave as a single shell with an averaged $e^\pm$ density and a common 
electron spectrum. This is not a good approximation as discussed in detail
below. In particular, at $R>\Rload$, only a small fraction of the 
blast-wave material is dominated by pairs, and its emission dramatically 
differs from the rest of swept-up material.

Most recently, Kumar \& Panaitescu (2004) studied $e^\pm$-loaded blast waves.
They focused on GRBs where the $\gamma$-ray front only partially overtakes 
the shock wave at $\Rload$, so that only a leading part of the front 
creates $e^\pm$ ahead of the shock, which reduces the number of shocked 
$e^\pm$. A dramatic effect appears in that situation, which was neglected 
in Kumar \& Panaitescu (2004): since the postshock $e^\pm$ still overlap 
with the $\gamma$-ray front, they are exposed to $0.1-1$~MeV photons (keV 
in the plasma frame) and quickly cooled by inverse Compton scattering. 
Most of the $e^\pm$ energy is then emitted by upscattering $0.1-1$~MeV 
photons to GeV-TeV band, and their optical synchrotron emission is 
suppressed (Beloborodov 2005).

Whether the $\gamma$-ray front still overlaps with the blast wave at a 
radius of interest depends on the front thickness (proportional to the
duration of the prompt GRB) and the blast-wave Lorentz factor 
$\Gamma$.\footnote{The $\gamma$-ray front can be emitted when the explosion
has a small radius, well before the blast wave forms, however, they may 
still overlap at large radii because the ejecta expands with almost speed 
of light. The $\gamma$-ray front is faster by a small 
$\delta v=c/2\Gamma^2\sim 10^{-5}c$ and completely overtakes the relativistic 
blast wave at time $R/c\simgt\Delta/\delta v$.}
So, two qualitatively different regimes of early afterglows are possible: 

\noindent
1. --- ``Long-burst'' (or ``thick-shell'') regime where the early blast 
wave overlaps with the prompt GRB radiation. Then a strong GeV-TeV flash 
should be produced, and the early optical emission is suppressed.

\noindent
2. --- ``Short-burst'' (or ``thin-shell'') regime where the prompt 
$\gamma$-rays early overtake the external shock wave, and $e^\pm$ creation 
and Compton cooling take place {\it ahead} of the shock. Then the postshock 
plasma is not exposed to the prompt radiation and not Compton cooled. 
A bright optical emission can be expected in this situation.

In the present paper, we focus on the short-burst regime. GRBs that satisfy 
this condition have short durations $t_b$ and/or modest Lorentz factors of 
the blast wave $\Gamma$ (B02),
\be\label{dur}
  t_b<10\,\left(\frac{\Gamma}{100}\right)^{-2}
    \left(\frac{\Racc}{10^{16}\rm cm}\right)\,\frac{(1+z)}{2}\;\;{\rm s},
\ee 
where $\Racc=5^{-1/2}\Rload$ is a characteristic radius where most of the 
optical-emitting pairs are created, and $z$ is the cosmological redshift 
of the burst. We develop a suitable Lagrangian formalism that 
describes the synchrotron emission of pair-loaded blast waves and calculate 
the expected light curve and spectrum of the early afterglow. We find two 
distinct emission components produced at $R>\Rload$: (1) the relict 
$e^\pm$ radiation component dominating the early emission from IR to 
soft X-rays, and (2) the recently shocked pair-free component that gives
a standard afterglow emission initially peaking in X-rays and later 
evolving to softer bands. 

In \S~2 we briefly describe the pair creation process and the formation of 
blast wave in the pair-loaded medium behind the $\gamma$-ray front 
(details are found in B02). In \S~3 we formulate the emission problem for 
$e^\pm$-loaded blast waves and develop their Lagrangian description. 
Numerically calculated examples of $e^\pm$-loaded afterglows are given in 
\S~4.  

In \S~5 we show that the $e^\pm$ component of afterglow emission is 
described by a simple formula and practically independent of the details of 
the shock wave physics. The case of a uniform ambient medium is elaborated
in \S~6. In the present paper, we focus on explosions in media of modest 
density $n_0=0.1-10^3$~cm$^{-3}$. Explosions in a high-density wind from a 
Wolf-Rayet progenitor will be considered in detail elsewhere. In that case, 
the $e^\pm$-loading has a much stronger effect on the afterglow emission 
(B02).

There are two shock fronts in a blast wave: forward and reverse.
The reverse shock emits one more component of the early 
afterglow, which depends on the nature of the GRB ejecta. 
Differences between emissions from the reverse shock 
and the pair-loaded forward shock are discussed in \S~7.
The differences are significant and may allow one to distinguish 
observationally the two emission mechanisms when the {\it Swift} 
satellite provides the early afterglow data.


\section{Pair loading by the $\gamma$-ray front}

A medium overtaken by a front of collimated $\gamma$-rays is inevitably
$e^\pm$-loaded. This happens because some $\gamma$-rays Compton scatter 
off the medium and get absorbed by the primary collimated radiation via 
reaction $\gamma+\gamma\rightarrow e^++e^-$. 

The medium is optically thin, so only a tiny fraction of the GRB radiation 
front scatters and turns into $e^\pm$, however, the number of created 
$e^\pm$ {\it per ambient electron} can be very large, $n_\pm/n_0\gg 1$.
The column density of an expanding photon front scales with radius as 
$R^{-2}$; hence the number of scattered photons per ambient electron 
is decreasing as $R^{-2}$ and $n_\pm/n_0\gg 1$ at small $R$. The created 
$e^\pm$ do 
more scattering, which leads to an exponential runaway of pair creation. 
There is a sharp boundary $\Rload$ between the exponentially loaded $e^\pm$ 
region and the outer pair-free region. The pair creation and scattering 
of GRB radiation inside $\Rload$ is accompanied by momentum deposition and 
the medium is accelerated radially away from the center.

The pair loading factor $n_\pm/n_0$ and acceleration of an optically thin
medium in the $\gamma$-ray front do not depend on $n_0$ and can be 
calculated starting with just one 
ambient electron and one ambient proton to which the electron is coupled.
The $e$ and $p$ components of the medium move always together to maintain
neutrality of the plasma. Any momentum communicated to $e$ is immediately
shared with $p$, so that the effective mass of $e$ is $m_e+m_p$. 
As the $\gamma$-ray front passes through, the electron component is
enriched by a number of additional $e^\pm$ (then the mass per lepton is 
reduced) and altogether they acquire a Lorentz factor $\gamma$. 
This transformation is quick: it takes time $\sim(\Delta/c\gamma^2)$ in 
the fixed lab frame, during which the $\gamma$-ray front overtakes a given 
ambient electron.\footnote{Velocity difference between the radiation 
front and the accelerated medium is $(1-\beta)c\approx c/2\gamma^2$.}
The thickness of the $\gamma$-ray front $\Delta$ is related to 
the observed duration of the prompt GRB $t_b$ by $t_b=(1+z)\Delta/c$. 

A schematic explosion picture is shown in Figure~1. The radiation front 
leads the forward shock by a small distance\footnote{We assume that the 
shock wave is approximately adiabatic. Then 
$\Gamma_{\rm shock}\approx\sqrt{2}\Gamma$, where $\Gamma$ is the Lorentz 
factor of the postshock material (which we hereafter call ``blast'' for
brevity), and
$l\approx R(1-\beta_{\rm shock})=R/2\Gamma^2_{\rm shock}=R/4\Gamma^2$.}
\be
  l\approx \frac{R}{4\Gamma^2}\ll R,
\ee
and changes the ambient medium just before it is shocked. The medium ahead 
of the blast wave (but already overtaken by the $\gamma$-ray front) is 
described by the lepton number per ambient proton $Z$ and Lorentz factor 
$\gamma$. $Z=1$ and $\gamma=1$ would correspond to a static pair-free medium.

\begin{figure}
\begin{center}
\includegraphics*[width=15cm]{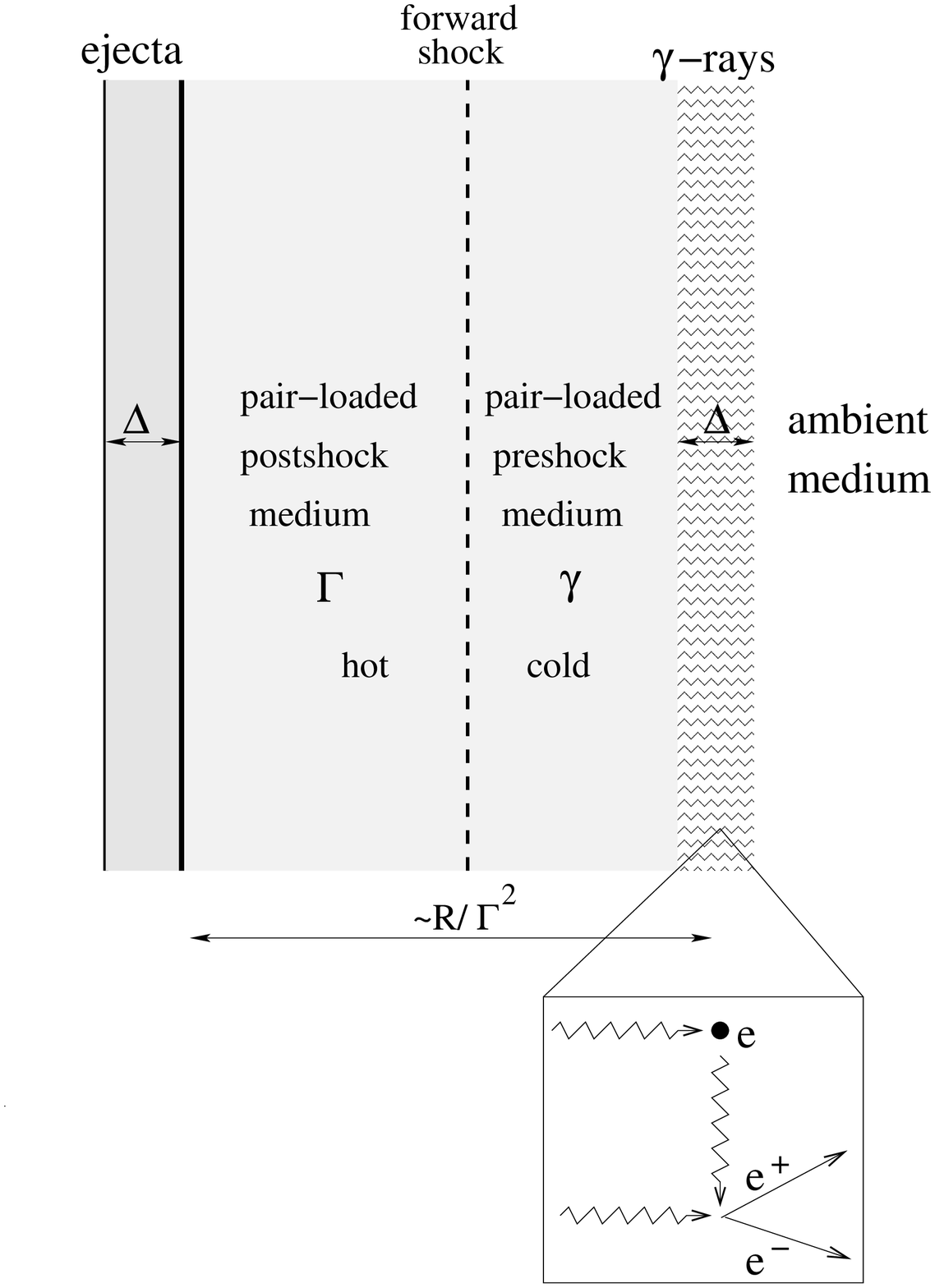}
\caption{ Early stage of a GRB blast wave ($R<\Rload$). The short-burst 
(or thin-shell) regime is assumed: $\Delta<R/4\Gamma^2$ (cf. eq.~1). The 
forward shock propagates in the $e^\pm$-loaded and preaccelerated medium 
left behind the $\gamma$-ray front. The pair-loading factor $Z$ and Lorentz 
factor $\gamma$ of the preshock medium depend on the current radius $R$ of 
the explosion (see Fig.~2). The inset schematically shows what happens in 
the $\gamma$-ray front: some of the $\gamma$-rays are scattered, lose their 
collimation and get absorbed by the collimated $\gamma$-rays, producing 
$e^\pm$. The created pairs are immediately Compton cooled in the 
$\gamma$-ray front (B02), and the preshock medium is relatively cold 
($kT<m_ec^2$). The pairs are heated/accelerated when the forward shock 
reaches them, and the postshock $e^\pm$ produce the broad-band synchrotron 
emission. The whole structure --- radiation front, $e^\pm$-loaded preshock 
medium, blast, and ejecta --- has a small thickness $\sim R/\Gamma^2$. 
}
\end{center}
\end{figure}

$Z$ and $\gamma$ are functions of only one parameter of the front, $\xi$, 
which we now define. Let $\dd E_\gamma/\dd S$ be the energy column density 
of the $\gamma$-ray front [erg/cm$^2$]. The GRB is likely beamed, yet it 
is convenient to define its isotropic equivalent 
$E_\gamma=4\pi R^2(\dd E_\gamma/\dd S)$. 
When an ambient electron is overtaken by the front, it scatters 
energy\footnote{Accurate calculation of scattering includes the 
Klein-Nishina correction to Thomson cross section $\sT$, however, in the 
definition of $\xi$, it is convenient to use $\sT$ which is independent of 
photon energy.} $e_{\rm sc}=\sT(\dd E_\gamma/\dd S)=E_\gamma\sT/4\pi R^2$.
The relevant dimensionless parameter is
\begin{equation}
\label{eq:xi}
  \xi=\frac{e_{\rm sc}}{m_ec^2}=65E_{\gamma,53}R_{16}^{-2}.
\end{equation}
$Z(\xi)$ and $\gamma(\xi)$ were found numerically in B02. 
They are calculated by solving the radiative transfer problem 
coupled to the dynamic problem of the medium acceleration. 
We briefly summarize the calculations here.

Two processes play important roles in the $\gamma$-ray transfer: Compton 
scattering $\gamma+e^\pm\rightarrow \gamma_{\rm sc}+e^\pm$ and photon-photon 
absorption $\gamma_{\rm sc}+\gamma\rightarrow e^++e^-$ (here symbol $\gamma$ 
stands for a primary photon and $\gamma_{\rm sc}$ for a scattered photon). 
The same processes determine the deposited momentum and acceleration of the 
medium. The medium remains optically thin, so only single scattering is of 
interest, and the process 
$\gamma_{\rm sc}+e^\pm\rightarrow\gamma_{\rm sc}+e^\pm$ is negligible. 
Nevertheless, there is a non-linearity in the problem because the scattering 
opacity of the medium is affected by $e^\pm$ creation and changes enormously 
across the front. 

The problem would be simpler if the scattered photons instantaneously 
converted to $e^\pm$ (Thompson \& Madau 2000; M\'esz\'aros, Ramirez-Ruiz, 
\& Rees 2001) --- then there would be no need to solve the radiative 
transfer. However, the bulk of scattered photons never convert to $e^\pm$ 
and escape. The formal absorption free-path of a scattered photon in the 
radiation front turns out to be $\lambda_{\gamma\gamma}\gg\Delta$, which 
means that a scattered photon has a small chance 
$\Delta/\lambda_{\gamma\gamma}$ to create an $e^\pm$ pair. 
The exponential runaway of pair loading occurs when the smallness of this 
chance is compensated by a large number of scattered photons per electron, 
which requires a small scattering free path of the electron in the radiation 
front, $\lambda\ll \Delta$.  The length of exponential $e^\pm$ loading is 
then $a=(\lambda\lambda_{\gamma\gamma})^{1/2}<\Delta$.
The characteristic loading radius $\Rload$ is defined by $a=\Delta$;
$\lambda\ll \Delta\ll \lambda_{\gamma\gamma}$ at this radius.

At small $R$, the $e^\pm$-loaded medium is accelerated to a significant 
Lorentz factor $\gamma$ (not to be confused with the photon symbol $\gamma$
in the reaction formulae). The photons scattered by the accelerating medium 
have the collimation angle $\delta\theta\approx \gamma^{-1}$ and their 
chances to convert to pairs become completely negligible (a smaller angle 
between the scattered and primary photons implies a higher threshold for 
reaction $\gamma_{\rm sc}+\gamma\rightarrow e^++e^-$ and a smaller absorption 
opacity seen by the scattered photons). Therefore, $e^\pm$ loading at small 
radii is made by photons scattered in a small leading portion of the front 
(where the medium has not yet acquired $\gamma\gg 1$) and propagated across 
the front.

The numerical solution to the transfer problem describes what exactly 
happens with the medium in the radiation front. The shell of ambient medium 
that is inside the front at a given moment of time has a certain velocity 
profile $\beta(\varpi)$ where $0<\varpi<\Delta$ is the distance from the 
leading edge of the front (``entrance''). The medium velocity $\beta$ 
increases from zero ($\gamma=1$) at the entrance $\varpi=0$ to its final 
value at the exit $\varpi=\Delta$. Lepton number
per proton, $Z$, also increases from $Z=1$ to its value behind the front.
The front structure is described by the same functions $\gamma(\xi)$ and
$Z(\xi)$ that describe the front evolution with radius if one substitutes
$\xi=\varpi/\lambda$ (B02). So, the front structure is described by the 
same functions at different times, i.e., it is self-similar. 

Since there is a gradient of the medium velocity inside the front ---
layers at larger $\varpi$ move faster --- one might think that caustics,
i.e., internal shocks can develop in the front. It never happens at radii
of interest. The radiation overtaking the medium dictates its velocity, 
and the medium has no time to develop a caustic because it quickly exits 
the front\footnote{The exit time $\Delta/c\gamma^2$ is much shorter than 
$R/c$ at $R>\Rgap$ (\S~2.1). The fact that no caustics develop in the front 
at $R>\Rgap$ formally follows from equation (12) in B02, which shows that 
density remains finite at all $\varpi$, while a caustic would correspond 
to infinite density. At an earlier stage, $R<\Rgap$, which is not considered 
in this paper, caustics do develop, see B02.} and 
is left behind with a uniform Lorentz factor $\gamma$ (the exit $\gamma$ 
gradually evolves on timescale $\sim R/c$ because $\xi=\Delta/\lambda$ 
evolves as $R^{-2}$ as the front expands). The medium left behind the 
radiation front is immediately picked up by the blast wave, on a timescale 
$\sim (R/c)(\Gamma/\gamma)^2\ll R/c$.

The exact numerical solution for $\gamma(\xi)$ and $Z(\xi)$ is well 
approximated by a simplified analytical model derived in B02 and summarized 
in the next section.

\subsection{Analytical description of $e^\pm$-loading}

The exact $Z(\xi)$ and $\gamma(\xi)$ are approximated by the following 
analytical formulae (see eqs.~49, 62, 63 in B02; $Z=n^*/n_0$ in B02 
notation),
\begin{equation}
\label{eq:Z}
   Z(\xi)=\left\{\begin{array}{ll}
    \frac{1}{2}[\exp(\xi/\xiload)+\exp(-\xi/\xiload)] & \xi<\xiacc, \\
    (\xi/\xiacc)^2\Zacc & \xiacc<\xi<3\xiacc, \\
  3(\xi/\xiacc)\Zacc & \xi>3\xiacc,\\
  \end{array}\right.
\end{equation}
\begin{equation}
  \label{eq:gam}
   \gamma(\xi)=
  \left\{\begin{array}{ll}
    1 & \xi<\xiacc, \\
    (\xi/\xiacc)^3 & \xiacc<\xi<3\xiacc, \\
  3\sqrt{3}(\xi/\xiacc)^{3/2} & \;\; \xi>3\xiacc,\\
  \end{array}\right.
\end{equation}
where $\xiacc=(5+\ln\mu_e)\xiload$, 
\be
\label{eq:Zacc}
  \Zacc=\frac{1}{2}\left[\exp\left(\frac{\xiacc}{\xiload}\right)
      +\exp\left(-\frac{\xiacc}{\xiload}\right)\right]=74\mu_e,
\ee
and $\mu_e$ is the electron mean molecular weight of the ambient 
medium ($\mu_e=1$ for hydrogen and $\mu_e=2$ for helium or heavy ions).
The numerical value of $\xiload=20-30$ depends on the precise spectrum of 
the $\gamma$-rays, however all the other relations remain the same.

The standard GRB spectrum is a broken power-law with a peak at 
$h\nu\approx m_ec^2$,
\be
\label{eq:F_gamma}
  F_\nu=F_1\left\{\begin{array}{ll}
    (h\nu/m_ec^2)^{-\alpha_1} & h\nu<m_ec^2, \\
    (h\nu/m_ec^2)^{-\alpha_2} & h\nu>m_ec^2. \\
  \end{array}\right.
\ee 
When observed from a redshift $z\approx 1$, the spectrum peaks at 
$h\nu\approx 250$~keV, as reported by BATSE observations (Preece et al. 2000).
$\xiload$ for such a spectrum was derived in B02,
\be
\label{eq:xiload_}
\xiload(\alpha_1,\alpha_2)=\frac{(\alpha_2-\alpha_1)}
    {(1-\alpha_1)(\alpha_2-1)}\left(\frac{\alpha_2-\alpha_1}
    {2\phi\epsilon_{\rm KN}^{\alpha_2-\alpha_1}}\right)^{1/2},
\ee
where $\epsilon_{\rm KN}\approx 0.4$ and 
$\phi=\frac{7}{12}\,2^{-\alpha_2}(1+\alpha_2)^{-5/3}$. 
In the examples below we fix $\alpha_1=0$, so that
\be
\label{eq:xiload}
   \xiload(\alpha_2)=\left(\frac{6}{7}\right)^{1/2}
   \frac{5^{\alpha_2/2}\alpha_2^{3/2}}{\alpha_2-1}
   \left(1+\alpha_2\right)^{5/6}.
\ee
A typical GRB has $\alpha_2\approx 1.5$, which gives $\xiload\approx 120$.

At sufficiently large radii, when the $\gamma$-ray front has $\xi<\xiload$, 
its effect on the medium is negligible: the medium remains almost static 
($\gamma\approx 1$) and $e^\pm$-free ($Z\approx 1$). When the front has 
$\xi>\xiload$, the runaway $e^\pm$ loading occurs. 
The number of loaded pairs depends exponentially on $\xi$ as long as 
$\xi<\xiacc$. At $\xi>\xiacc$, the front acts as a relativistic accelerator 
and the dependence of $\gamma$ and $Z$ on $\xi$ can be approximated by power 
laws (eqs.~\ref{eq:Z},\ref{eq:gam}). The slopes of $\gamma(\xi)$ and $Z(\xi)$ 
change at $\xi\approx 3\xiacc$ where $Z\approx 10^3$ and the mass of 
injected $e^\pm$ is comparable to that of the ambient ions. 
An interesting effect takes place at 
$\xi>\xigap\approx 3\times 10^3$: then $\gamma(\xi)$ exceeds the Lorentz 
factor of the ejecta $\Gej=10^2-10^3$. The radiation front with such a high 
$\xi$ pushes the external medium away from the fireball and opens a gap,
so that the fireball moves in a cavity cleared by its own radiation front.

The $\gamma$-ray front expands with time, and its $\xi$-parameter decreases
as $R^{-2}$. It starts at very high $\xi$ and then passes through $\xigap$, 
$\xiacc$, and $\xiload$ at radii $\Rgap$, $\Racc$, and $\Rload$, respectively,
\be
\label{eq:Rxi}
\Racc\approx 7\times 10^{15}\left(\frac{E_\gamma}{10^{53}\rm erg}\right)^{1/2}
  \cm,\qquad \Rload=(5+\ln \mu_e)^{1/2}\Racc, \qquad \Rgap\approx \Racc/3.
\ee

$Z(R)$ and $\gamma(R)$ of the preshock medium 
are shown in Figure~2 for $E_\gamma=10^{53}$~\erg. They do not depend
on the density of the ambient medium as long as the medium is optically
thin and are entirely determined by the parameters of the radiation front. 
The figure shows the exact $Z$ and $\gamma$ (cf. Fig.~1 and 3 in B02) 
and their analytical approximations (eqs.~\ref{eq:Z} and \ref{eq:gam}).

\begin{figure}
\begin{center}
\plotone{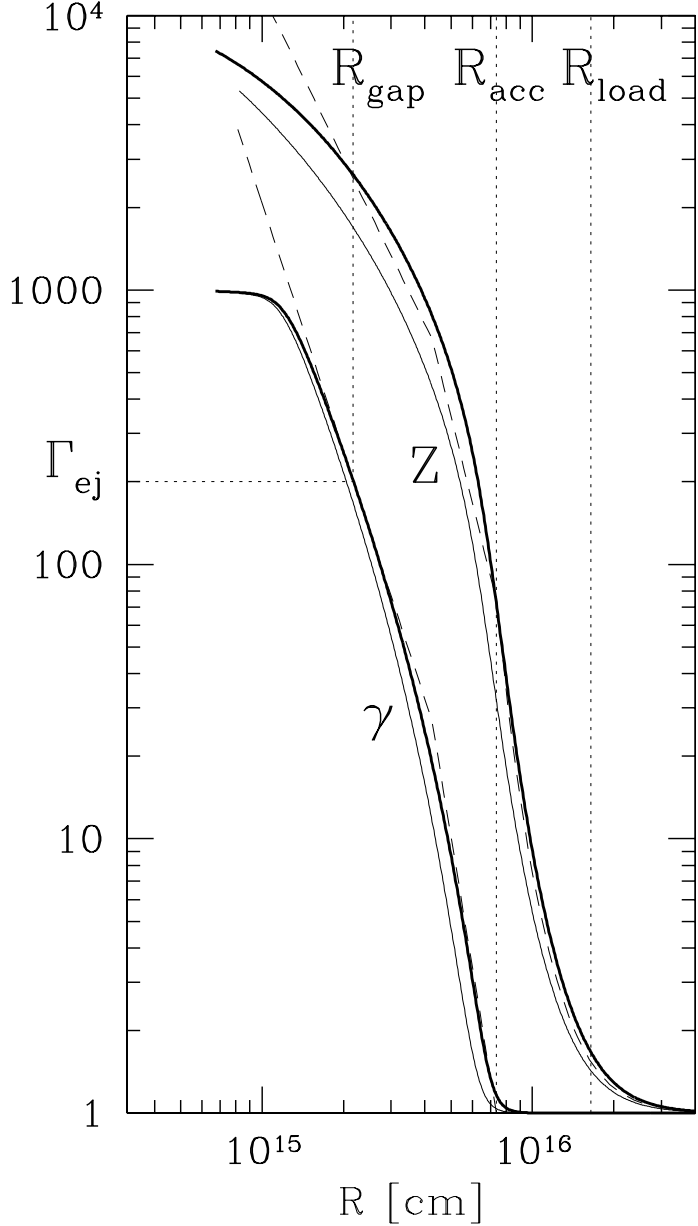}
\label{fig:front}
\end{center}
\caption{\small Pair-loading factor $Z$ and Lorentz factor $\gamma$ acquired by 
a medium at a radius $R$ when it is overtaken by a $\gamma$-ray front. 
The front has isotropic energy $E_\gamma=10^{53}$~erg. The exact numerical
results are shown by solid curves and the analytical approximation
(eqs.~\ref{eq:Z} and \ref{eq:gam}) by dashed curves.
The characteristic radii $\Rgap$, $\Racc$, and $\Rload$ are indicated 
by vertical dotted lines. The condition $\gamma=\Gej$ defines $\Rgap$; 
in this example $\Gej=200$ (shown by the horizontal dotted line).
$Z(R)$ and $\gamma(R)$ for different $E_\gamma$ are found
by simple re-scaling $R\rightarrow (E_\gamma/10^{53})^{1/2}R$.
The GRB spectrum assumed in the calculation has the typical high-energy 
slope $\alpha_2=1.5$. To illustrate the dependence on $\alpha_2$
we also show $Z(R)$ and $\gamma(R)$ for $\alpha_2=2$ (thin solid curves).
}
\end{figure}

Variations in GRB spectra have some effect on $Z(R)$ and $\gamma(R)$
because they affect $\xiload$. For instance, if $\alpha_2$ changes 
from 1.5 to 2 then $\xiload$ changes from 24 to 33 (eq.~\ref{eq:xiload}); 
the resulting changes in $Z(R)$ and $\gamma(R)$ are shown in Figure~2.

\subsection{Blast wave}

The prompt GRB emission can be produced quite early inside the fireball, 
preceding the development of the external blast wave at 
$R\sim 10^{15}-10^{17}$~cm. The millisecond variability observed in GRBs 
is often cited as an evidence for small radii of $\gamma$-ray emission, 
$R_\gamma\sim 10^{12}-10^{14}$~cm. Regardless the precise $R_\gamma$, the 
$\gamma$-rays are definitely emitted before the afterglow and then propagate 
ahead of the blast wave through the ambient medium.

The three characteristic radii $\Rgap$, $\Racc$, and $\Rload$ define four 
stages of the GRB explosion:
\medskip

{\bf I.} $R<\Rgap$. The fireball moves in a cavity cleared by its 
radiation front. The $e^\pm$-rich external medium ($Z>10^3$) surfs ahead 
with $\gamma>\Gej$.

{\bf II.} $\Rgap<R<\Racc$. The external blast wave has formed and the
fireball sweeps up the $e^\pm$-rich medium 
($Z\simgt 10^2$) that moves with $1\ll\gamma<\Gej$.

{\bf III.} $\Racc<R<\Rload$. The fireball sweeps up the 
static medium ($\gamma\approx 1$) dominated by $e^\pm$ ($Z\gg 1$).

{\bf IV.} $R>\Rload$. The fireball sweeps up the static pair-free medium.
\medskip

The blast wave develops at $R>\Rgap$, with its forward and reverse shocks.
The preshock external medium is $e^\pm$-rich and
moving relativistically, which affects the jump conditions and the state of 
the postshock plasma. At $R=\Rgap$ the blast wave
gently begins to sweep up the preaccelerated medium with a small relative
Lorentz factor $\Grel\approx 1$ ($\Gej\approx\gamma$). With increasing 
$R>\Rgap$, $\gamma$ falls off quickly and approaches unity at $R=\Racc$ as
$\gamma=(R/\Racc)^{-6}$. Thus, the fireball suddenly ``learns'' that there 
is an interesting amount of slow material in its way and hits it with a 
large $\Grel$. This resembles a collision with a wall and results in a 
sharp rise of the afterglow at $R\approx\Racc$.

The explosion picture is modified if the $\gamma$-ray front is created by 
the shock wave itself (external model of the prompt GRB) rather than by 
internal dissipation in the fireball. Then a self-consistent blast-wave 
model exists without the gap. A sufficiently dense medium can complicate 
that picture if it leads to an ``electro-magnetic catastrophe'' in the 
forward shock (Stern 2003) --- a runaway production of $\gamma$-rays, 
which may result in transient gap openings.

The uncertainties in the mechanism of the prompt GRB play little role for 
the afterglow model developed below as we focus on the stage when the 
process of $\gamma$-ray production is over. We consider $R>\Rgap$, and the
blast wave at these radii has practically no memory of the gap 
opening. We need only to know the parameters of the emitted $\gamma$-ray 
front (eq.~\ref{eq:F_gamma}) which are taken from observations. 
Therefore, the following calculations apply to both internal and external 
models of the prompt GRB emission.


\section{Afterglow calculation}

The production of afterglows without the $e^\pm$-loading effects, 
i.e. with $Z=1$ and $\gamma=1$, was studied previously in detail.
The electrons were assumed to receive a significant fraction 
$\epsilon_e\sim 0.1$ of the shock energy (the remaining fraction 
$1-\epsilon_e$ is carried by the postshock ions). This leads to a high 
energy per electron, and the early high-$\Gamma$ blast wave emits 
hard X-rays in the fast-cooling regime: the electrons immediately radiate 
the received energy, practically at the same radius where they are shocked. 

The $e^\pm$ loading changes the picture: a lot of leptons now share the 
energy received from the shock. As a result, the mean energy per postshock 
$e^\pm$ is reduced by orders of magnitude and $e^\pm$ emit much softer 
radiation. Furthermore, the postshock $e^\pm$ can be in the slow-cooling 
regime so that their emission remains significant at $R>\Rload$ where 
no new pairs are added to the blast wave.

\subsection{Formulation of the problem}

We focus on a sufficiently early stage when $\Gamma>\theta^{-1}$ where 
$\theta$ is the opening angle of the explosion; then the expanding jet 
behaves like a portion of a spherically-symmetric explosion and we can 
neglect the beaming effects. Three further technical assumptions simplify 
our calculations:

1. --- The shocked ambient material
is assumed to have a common Lorentz factor $\Gamma$ 
(relative motions in the postshock material are subsonic and 
the assumption of common $\Gamma$ is good within a factor of $1/\sqrt{3}$).

2. -- The postshock material is assumed to be in pressure 
equilibrium, i.e. $P=const$ throughout this material.

3. --- When calculating the observed luminosity we neglect the finite 
thickness of the blast wave 
($=$ the distance between the contact discontinuity and the forward 
shock). The blast-wave radiation is then characterized by its total 
instantaneous luminosity $L_\nu$ emitted at a current radius. 
Inclusion of time 
delays between photons from different subshells of the blast material 
would change the result by a factor of a few. A precise calculation of 
this factor would require to relax assumptions 1 and 2.

These assumptions will allow us to derive the instantaneous 
luminosity $L_\nu$ at a given frequency $\nuobs$. We are 
interested in the soft emission here, from IR to soft X-rays,
and especially optical, $\nuobs\approx 0.5\times 10^{15}$~Hz.

Consider the moment of time when the blast wave reaches a given radius $\tR$.
The total postshock ambient mass at this moment is\footnote{Hereafter 
quantities taken at the current radius $\tR$ are marked by tilde to 
distinguish them from the corresponding quantities at radius $R(m)$ where 
shell $(m,m+\delta m)$ was shocked.} 
\be
  \tm(\tR)=\int_0^{\tR} 4\pi R^2\rho_0(R)\dd R.
\ee
Each subshell $\delta m$ of this material has its own history.
It used to be a shell of ambient medium located at a radius $R<\tR$, 
which was loaded with $e^\pm$ pairs and preaccelerated by the prompt 
$\gamma$-rays, then immediately shocked and picked up by the 
relativistic blast wave. The radius where all that happened
is related to $m$ by equation
\be
   m=\int_0^{R(m)} 4\pi R^2\rho_0(R)\dd R.
\ee
Then $\delta m$ cooled radiatively and 
adiabatically as the blast wave expanded to the present radius $\tR$.

We need to evaluate the contribution of each $\delta m$ to the current 
luminosity $L_\nu(\tR)$ and therefore have to resolve the blast wave 
material in its (Lagrangian) mass coordinate $0<m<\tm$, even though
we do not resolve it in radius and assume that all $\delta m$ are located
at the same $\tR$. $m\rightarrow 0$ is the material 
which was swept up first (at small radii),
and $m\rightarrow \tm$ is the currently shocked material.
$L_\nu$ is the sum of current emissions $\delta L_\nu$ from all $\delta m$.  

In afterglow models without $e^\pm$ loading
the emission from $m\ll \tm$ is negligible and the
instantaneous luminosity is dominated by the
recently shocked mass shell $m\simlt \tm$. 
A special feature of pair-loaded blast waves is that their $L_\nu$ 
peaks at $m\ll \tm$. Therefore an accurate integration over $m$ is needed.

To calculate $\delta L_\nu$ from each $\delta m$ we need to know the 
current $e^\pm$ spectrum $\tf(\gamma_e)$ and magnetic field $\tB$ in 
$\delta m$.

\subsection{Magnetic field}

The magnetic field in shell $\delta m$ in the blast wave of radius $\tR$ 
is given in terms of the customary equipartition parameter $\tepB(m)<1$,
\be
\label{eq:B}
   \tilde{U}_B(m)=\frac{\tB^2(m)}{8\pi}=\tepB(m)\tU,
\ee
where $\tU=3\tP$ is the current energy density in the blast wave and 
$\tP$ is its pressure.
$\tP$ and $\tU$ are assumed to be constant throughout the postshock material
in our simplified model. They are functions of $\tR$ only and related to 
the ambient density by the jump condition at the shock front,
\be
\label{eq:U}
   \tU=3\tP=4\trho_0 c^2\tGrel\tG,
\ee
where
\be
\label{eq:Grel}
 \tGrel=\frac{\tG}{\tg(1+\tb)}  \qquad ({\rm when}\;\, \tg\ll\tG)
\ee
is the current Lorentz factor of the preshock medium relative to the blast 
wave, and $\tg=(1-\tb^2)^{-1/2}$ is the Lorentz factor of the preshock 
medium in the lab frame. The jump condition~(\ref{eq:U}) takes into account
the compression of the ambient medium preaccelerated to a Lorentz factor 
$\tg$ (Madau \& Thompson 2000; B02). Thus, the magnetic field can be 
written as
\be
\label{eq:tB}
  \tB=\tG\left[\frac{32\pi\,\tepB\,\trho_0 c^2}{\tg(1+\tb)}\right]^{1/2}
  =0.39\,\tG\left[\frac{\tepB \tn_0\mu_e}{\tg(1+\tb)}\right]^{1/2} \,{\rm G},
\ee
where $\tn_0$ is expressed in cm$^{-3}$.

$\tepB(m)$ can change with $\tR$:
the magnetic field evolves as $\delta m$ expands from the radius 
$R$ where it was shocked to the current radius $\tR$. The toroidal field 
component is dominant in the expanding shocked plasma and, if the magnetic 
flux is conserved, $\tB$ evolves as
\be
\label{eq:fc}
  \frac{\tB(m)}{B(m)}=\frac{\trho}{\rho}\,\frac{\tR}{R}
       =\left(\frac{\tP}{P}\right)^{3/4}\,\frac{\tR}{R},
\ee
where $\rho$ is the proper mass density of baryons and we have used
the adiabatic equation of state $P\propto \rho^{4/3}$. Using 
equation~(\ref{eq:U}), we can rewrite the flux-conservation condition as
\be
\label{eq:tepB}
  \frac{\tB(m)}{B(m)}=\left(\frac{\tG\tGrel\trho_0}{\Gamma\Grel\rho_0}
                      \right)^{3/4} \,\frac{\tR}{R}, \qquad
   \frac{\tepB}{\epB}=\left(\frac{\tG\tGrel\trho_0}{\Gamma\Grel\rho_0}
                      \right)^{1/2}\,\left(\frac{\tR}{R}\right)^2.
\ee
It shows that $\tepB$ grows downstream of the shock. This growth, however, 
cannot proceed beyond equipartition: $\tepB$ then saturates near unity.

If the flux is not conserved (the magnetic field may reconnect/dissipate in
the blast) the evolution of postshock $\tB$ is different. Therefore, we will 
not specify $\tepB$ in the calculations until the last step when we give 
examples. Then, besides equation~(\ref{eq:tepB}) we will also consider the 
case of $\tepB=\epB=const$, which is a second reasonable prescription for 
the magnetic field evolution.

\subsection{Distribution function of electrons/positrons}

The customary phenomenological shock model of GRBs assumes that the 
electrons (and positrons) are impulsively accelerated at the shock front
with a power-law spectrum, 
\begin{eqnarray}
\label{eq:f}
   f(\gamma_e)=
  \left\{\begin{array}{ll}
    0, & \gamma_e<\gamma_m, \\
   K\left(\gamma_e/\gamma_m\right)^{-p}, & \gamma_e>\gamma_m.\\
  \end{array}\right.
\end{eqnarray}
This initial spectrum is injected in $\delta m$ at the radius $R(m)$ where 
$\delta m$ is shocked; the Lorentz factor $\gamma_e$ of the accelerated 
leptons is measured in the fluid frame.
The total injected energy of nonthermal $e^\pm$ in this frame is  
\be
\label{eq:dE}
  \delta E=\int \gamma_e m_ec^2f(\gamma_e)\,\dd\gamma_e
          =\epe\Grel\,\delta m\, c^2,
\ee
and the total number of nonthermal leptons in $\delta m$ is
\be
\label{eq:dN}
 \delta N=\int f(\gamma_e)\,\dd\gamma_e
         =\left(\frac{\delta m}{\mu_e m_p}\right)Z.
\ee
Here $p>2$ and $\epe<1$
are phenomenological parameters of the electron acceleration.

From equations~(\ref{eq:dE}) and (\ref{eq:dN}) one finds $\gamma_m$ and $K$,
\be
\label{eq:gm_}
  \gamma_m=\frac{\Grel}{(Z/\mu_e)}\frac{\epe(p-2)m_p}{(p-1)m_e}, 
\ee
\be
  K=(p-1)\frac{\delta N}{\gamma_m}.
\ee

When a shocked $\delta m$ expands from $R$ to $\tR$ the injected nonthermal 
spectrum is modified by two effects: 

1. --- Adiabatic cooling shifts the whole $e^\pm$ distribution as 
$\gamma_e\propto P^{1/4}$, retaining the power-law shape. 
The minimum Lorentz factor of the nonthermal spectrum, $\gamma_m$, changes 
by the factor
\be
\label{eq:A}
  A=\left[\frac{\tP}{P}\right]^{1/4}
                =\left[\frac{\trho_0\tGrel\tG}
                {\rho_0\Grel\Gamma}\right]^{1/4},
\ee
\be
\label{eq:tgm}
   \tg_m(m,\tR)=A\,\gamma_m(m,R).
\ee
and the normalization $K$ of the spectrum changes as $\tK=K/A$.

2. --- Radiative cooling cuts off the $e^\pm$ distribution at high 
$\gamma_e$. In most of the models considered below, the exact position  
of the cutoff is not important because we focus on the low-frequency 
radiation. The cutoff can be estimated as follows.

The radiative cooling of $\delta m$ can peak at any radius $R^\prime$ 
between $R$ and $\tR$, and then the cutoff is shaped at this radius. 
Electrons with Lorentz factor $\gamma_e$ are cooled  with a rate 
(assuming isotropic pitch-angle distribution),
\be
  \dot{\gamma}_e=-\frac{4}{3}\frac{\sT}{m_e c}
    \left(\frac{{B^\prime}^2}{8\pi}+U_s^\prime\right)\gamma_e^2,
\ee
where $B^\prime=B(m,R^\prime)$ and $U_s^\prime=U_s(m,R^\prime)$ is the 
energy density of soft (synchrotron) radiation in $\delta m$ at radius 
$R^\prime$. We assume that the bulk of synchrotron radiation in the 
fluid frame satisfies $h\nufluid<m_ec^2/\gamma_e$ and scatters off $e^\pm$ 
with Thomson cross section. Using equations~(\ref{eq:B},\ref{eq:U}) we get
\be
 \dot{\gamma}_e=-\frac{16}{3}\frac{\rho_0^\prime}{m_e}\sT c\epB^\prime
                        \Grel^\prime\Gamma^\prime(1+C^\prime)\gamma_e^2. 
\ee
Here $C^\prime=U_s^\prime/U_B^\prime$ is the relative contribution of 
inverse Compton scattering to the cooling rate. 

The characteristic cooling Lorentz factor $\gamma_c^\prime$ 
at $R^\prime$ is defined by the condition that the cooling timescale 
$\gamma_e/|\dot{\gamma}_e|$ equals the expansion timescale
$(R^\prime-R)/\Gamma^\prime c$. This yields
\be
\label{eq:gcprime}
  \gamma_c^\prime(m)=\frac{3m_e}{16\epB^\prime\Grel^\prime\sT(1+C^\prime) 
                                 (R^\prime-R)\rho_0^\prime}.
\ee
Note that $C^\prime$ depends on the radiated energy $U_s^\prime$, which 
in turn depends on $\gamma_c^\prime$; therefore equation~(\ref{eq:gcprime})
is implicit. It can be solved as follows. The total radiation density in 
$\delta m$ is comparable to the lost $e^\pm$ energy:
$\Urad^\prime\approx (\gamma_c^\prime/\gamma_m^\prime)^{2-p}\epe U^\prime$ 
if $\gamma_c^\prime>\gamma_m^\prime$ and $\Urad^\prime\approx\epe U^\prime$
otherwise. Its synchrotron fraction is 
$U_s^\prime/\Urad^\prime=U_B^\prime/(U_s^\prime+U_B^\prime)$.
This allows one to express $C^\prime=U_s^\prime/U_B^\prime$ in terms of 
$\gamma_c^\prime$,
\be
\label{eq:C}
   C^\prime(C^\prime+1)=\frac{\epe}{\epB^\prime}\,
\left\{\begin{array}{ll}
    1 & \gamma_c^\prime<\gamma_m^\prime \\
 (\gamma_c^\prime/\gamma_m^\prime)^{2-p} & \gamma_c^\prime>\gamma_m^\prime \\
  \end{array}\right.
\ee
Now we have two equations~(\ref{eq:gcprime}) and (\ref{eq:C}) 
which can be solved for $C^\prime$ and $\gamma_c^\prime$.

This estimate assumes that $U_s^\prime(m)$ is produced locally at a given
$m$ and neglects the transport of synchrotron radiation across the blast 
wave. Inclusion of transport further complicates the 
calculation of $C^\prime(m)$ and $\gamma_c^\prime(m)$. In this paper, 
we avoid models with significant Compton cooling and use the simple 
estimate of $\gamma_c^\prime$ (eq.~\ref{eq:gcprime}) with $C^\prime=0$. 
The consistency of this estimate requires 
$C^\prime(\gamma_c^\prime)\simlt 1$, which can be checked using 
equation~(\ref{eq:C}). An estimate of Compton cooling that includes
the radiation transport will be given in \S~6.3. 

The $e^\pm$ distribution cutoff $\tg_c$ in $\delta m$ at the current radius 
$\tR$ is 
\be
\label{eq:tgc}
   \tg_c=\gamma_c^\prime\, A^\prime,
  \qquad A^\prime=\left[\frac{\trho_0\tGrel\tG}
   {\rho_0^\prime\Grel^\prime\Gamma^\prime}\right]^{1/4}.
\ee
It is shaped at the radius $R^\prime$ where $(\gamma_c^\prime\,A^\prime)$
is minimum with $\gamma_c^\prime$ calculated as described above, and its 
evolution from $R^\prime$ to $\tR$ is determined by the adiabatic cooling 
factor $A^\prime$. 

If $\tg_m<\tg_c$ (slow-cooling regime), the current nonthermal $e^\pm$ 
distribution in $\delta m$ is a power-law with slope $p$ and normalization 
$\tK$, extending from $\tg_m$ to $\tg_c$. If $\tg_c<\tg_m$ (fast-cooling 
regime), all nonthermal leptons $\delta N$ pile up near a single Lorentz 
factor $\gamma_e\approx \tg_c$. In any case, the number of nonthermal 
$e^\pm$ in $\delta m$ is given by equation~(\ref{eq:dN}), so their 
distribution function at $\tR$ can be written as
\begin{eqnarray}
\label{eq:tf}
   \tf(\gamma_e)=\delta N\,\times
  \left\{\begin{array}{ll}
\frac{(p-1)}{\tg_m}\left(\gamma_e/\tg_m\right)^{-p}
   \,H(\gamma_e-\tg_m)\,H(\tg_c-\gamma_e), & \tg_m<\tg_c.\\
  \delta(\gamma_e-\tg_c),  & \tg_m>\tg_c, \\
  \end{array}\right.
\end{eqnarray}
where $H(...)$ is Heaviside step function and $\delta(...)$ is 
Dirac $\delta$-function.

\subsection{Self-absorption}

Self-absorption can affect synchrotron emission from $e^\pm$ with low 
$\gamma_e$. The low-$\gamma_e$ population is created at the very beginning 
of pair-loading $(R<\Racc)$ when the preacceleration $\gamma$ is significant, 
the forward shock is relatively mild ($\Grel\ll\Gamma)$, and the 
pair-loading factor $Z$ is comparable to $10^3$ (cf. eq.~\ref{eq:gm_}). 
The low-energy postshock $e^\pm$ are slowly cooling and produce low-frequency 
radiation that is self-absorbed.

Self-absorption is, however, not important for blast-wave radiation with 
frequencies $\nu\simgt 10^{12}$~Hz that we consider thereafter. This
radiation is dominated by the blast-wave material with sufficiently high 
$\gamma_m$ so that self-absorption can be neglected for all 
$\gamma_e\geq\gamma_m$. The condition of small self-absorption can be 
written as $U_m < \gamma_m m_ec^2 (\nu/c)^3$, where $U_m$ is the density 
of synchrotron radiation produced by $e^\pm$ with $\gamma_e=\gamma_m$, 
and we assume thereafter
\be
  U_m\sim \gamma_m^2 U_B\sT\,\frac{R}{\Gamma}<\gamma_m m_ec^2
                                              \left(\frac{\nu}{c}\right)^3.
\ee
This roughly corresponds to $\nu>10^{12}$~Hz.

\subsection{Synchrotron luminosity}

Given the current magnetic field $\tB$ and nonthermal $e^\pm$ spectrum 
$\tf(\gamma_e)$ in $\delta m$, it is straightforward to evaluate its 
contribution $\delta L_\nu$ to the instantaneous luminosity of the blast 
wave. 

The synchrotron spectrum of $e^\pm$ with Lorentz factors $\gamma_e$
in the fluid frame peaks at the frequency (assuming isotropic pitch-angle
distribution) 
\be
\label{eq:nuf}
   \nufluid\approx 0.15\frac{e\tB}{m_ec}\,\gamma_e^2.
\ee
The intensity-weighted Doppler shift to the lab frame is given by
\be 
\label{eq:nu}
  \nu=\frac{4}{3}\,\tG\nufluid,
\ee
and the corresponding observed frequency is 
\be
\label{eq:nuobs}
  \nuobs=\frac{\nu}{1+z},
\ee
where $z$ is the cosmological redshift of the burst. 
From equation~(\ref{eq:nuf}) we find $\gamma_e$ of $e^\pm$ whose 
synchrotron spectrum peaks at a given $\nu$. We denote this characteristic 
Lorentz factor by $\tg_\nu$,
\be
\label{eq:tgnu}
  \tg_\nu\equiv\gamma_\nu(m,\tR)
   \approx\left[\frac{5\,\nu\, m_ec}{\tG e\tB}\right]^{1/2}.
\ee

The emitted synchrotron power by the $\tg_\nu$ electron is 
$\dot{E}_s=\sT c \tB^2\tg_\nu^2/6\pi$ and the synchrotron luminosity 
from an $e^\pm$ population with distribution $\dd N/\dd\gamma_e$
is approximately given by
\be
  \nu L_\nu = \frac{\dd L}{\dd \log\nu}
            = \frac{\dd L}{2\,\dd\log\tg_\nu}
 =\frac{1}{2}\left(\gamma_e\frac{\dd N}{\dd\gamma_e}\right)_{\gamma_e=\tg_\nu}
   \,\dot{E}_s(\gamma_\nu).
\ee
The $e^\pm$ population in $\delta m$ peaks near $\tg_m$ if
$\tg_c>\tg_m$ (slow-cooling regime) and near $\tg_c$ if
$\tg_c<\tg_m$ (fast-cooling regime). In the first case, the synchrotron 
luminosity of $\delta m$ can be written as
\begin{eqnarray}
\label{eq:dLslow}
  \delta\left(\nu L_\nu\right)
                   =\delta\left(\nufluid L_{\nu_{\rm fluid}}\right)
                   =\dLmax\times\left\{\begin{array}{ll}
    (\tg_\nu/\tg_m)^{2/3} & \tg_\nu<\tg_m<\tg_c, \\
    (\tg_\nu/\tg_m)^{1-p} & \tg_m<\tg_\nu<\tg_c, \\
    0 & \tg_m<\tg_c<\tg_\nu, \\
  \end{array}\right.
\end{eqnarray}
where 
\be
\label{eq:dLmax}
   \dLmax\approx \frac{\sT c\tB^2}{12\pi}\,\tg_\nu^2\,\delta N
    =\frac{5}{12\pi}\,\frac{m_ec^2\,\sT}{e}\,\frac{\tB}{\tG}\,\nu\,\delta N.
\ee
Note that $\delta(\nu L_\nu)$ is originally defined as energy emitted per 
unit time in the fixed lab frame, however, we calculate it in the fluid 
frame using the Lorentz invariancy of $\nu L_\nu$. 
$\dLmax$ is a maximum luminosity that would be emitted at $\nu$ at the most 
favorable condition $\tg_m=\tg_\nu$ [in this case $2\delta(\nu L_\nu$) 
equals the energy loss rate of the dominant $e^\pm$ population with 
$\gamma_e\sim \gamma_m$]. Similarly, in the fast-cooling regime, we have
\begin{eqnarray}
\label{eq:dLfast}
  \delta(\nu L_\nu)=\dLmax\times
  \left\{\begin{array}{ll}
    (\tg_\nu/\tg_c)^{2/3} & \tg_\nu<\tg_c<\tg_m, \\
    0 & \tg_\nu>\tg_c. \\
  \end{array}\right.
\end{eqnarray}
The nonthermal $e^\pm$ with Lorentz factors $\gamma_e=\tg_m$ and 
$\gamma_e=\tg_c$ emit radiation at frequencies
\be
\label{eq:tnu_mc}
   \tnu_m=0.2\tG\frac{e\tB}{m_ec}\,\tg_m^2,   \qquad
   \tnu_c=0.2\tG\frac{e\tB}{m_ec}\,\tg_c^2, 
\label{eq:tnuc}
\ee
and the spectral luminosity $\delta L_\nu/\delta m$ 
at a Lagrangian coordinate $m$ can be written as
\be
\label{eq:dL_nu}
  \frac{\delta L_\nu}{\delta m}=\frac{\dLmax_\nu}{\delta m}\times
  \left\{\begin{array}{ll}
    (\nu/\tnu_m)^{1/3} & \nu<\tnu_m<\tnu_c, \\
    (\nu/\tnu_m)^{(1-p)/2} & \tnu_m<\nu<\tnu_c, \\
    (\nu/\tnu_c)^{1/3} & \nu<\tnu_c<\tnu_m, \\
    0 & \nu>\tnu_c.
  \end{array}\right.
\ee
An explicit formula for $\dLmax_\nu=\dLmax/\nu$ is found using 
equation~(\ref{eq:dN}) for $\delta N$ and equation~(\ref{eq:tB}) for $\tB$,
\beq
\nonumber
  \frac{\dLmax_\nu}{\delta m } &=& \frac{5}{12\pi}\,\frac{m_ec^2\,\sT}{e}\,
                 \frac{\tB}{\tG}\,\frac{\delta N}{\delta m} 
  = c^2\left[32\pi\,\frac{m_e}{\mu_e m_p}\,
    \frac{\tepB\tn_0 r_e^3}{\tg(1+\tb)}\right]^{1/2} Z \\
 &\approx & 30\,\left[\frac{\tepB\tn_0}{\mu_e\tg(1+\tb)}\right]^{1/2} Z,
\label{eq:Lmax}
\eeq
where $r_e=e^2/m_ec^2\approx 2.81\times 10^{-13}$cm is the classical
electron radius. Note that $\dLmax_\nu/\delta m$ does not depend on $\nu$.

The total instantaneous luminosity of the blast wave is found by
integrating over $m$,
\be
\label{eq:Ltot}
   L_\nu(\tR) =\int_0^{\tm}\frac{\delta L_\nu}{\delta m}\,\delta m.
\ee

\subsection{Observed flux}

Observer at a distance $D$ much smaller than the Hubble 
scale would measure the spectral luminosity (e.g. Rybicki \& Lightman 1979),
\be
  \Lobs_\nu(\tR)=\frac{4}{3}\,\tG^2\, L_\nu,
\ee 
and the spectral flux $\Fnuobs=\Lobs_\nu/4\pi D^2$.
For cosmologically large $D$, the flux formula is modified by two effects: 
$D$ is replaced by the luminosity distance and the observed frequency
of radiation is redshifted, $\nuobs=(1+z)^{-1}\nu$,
\be
\label{eq:F}
  \Fnuobs= \frac{\nu L_\nu^{\rm obs}}{4\pi\nuobs D^2}
       =\frac{\tG^2(1+z)}{3\pi D^2}\,L_\nu,
\ee  
where
\be
\label{eq:D}
    D=\frac{2c}{H_0}\left(1+z-\sqrt{1+z}\right)
     \approx 2.6\times 10^{28}\left(1+z-\sqrt{1+z}\right) {\rm ~cm},
\ee
and $H_0\approx 70$~km/s/Mpc is Hubble constant.
The observed optical magnitude in V or R band is related to $\Fnuobs$ by
\be
  m_V=8.873-2.5\lg\Fnuobs,  \qquad 
  m_R=8.645-2.5\lg\Fnuobs,
\ee
where $\Fnuobs$ is in units of ${\rm Jy}=10^{-23}$~erg~cm$^{-2}$.

The flux $\Fnuobs(\tR)$ is received at the observer time
\be
\label{eq:tobs}
  \tobs(\tR)\approx \frac{\tR}{2\tG^2 c}\,(1+z)
\ee
after the arrival of first photons (prompt $\gamma$-rays) from the 
explosion. More exactly, it is received during a time interval 
$\Delta \tobs\sim (\tR/2\tG^2c)(1+z)$ because of the spherical curvature and 
finite thickness of the blast wave, and equation~(\ref{eq:tobs}) gives only
an approximate arrival time.
The numerical factor in this equation can vary around 1/2 by a factor 
$\sim 2$, depending on the blast wave dynamics $\tG(\tR)$.
We hereafter use the approximate equation~({\ref{eq:tobs}) with the fixed
factor $1/2$. This degree of accuracy is consistent with our approximate
treatment of the blast wave as a constant-pressure shell.


\section{Numerical examples}

It is straightforward to calculate the afterglow emission $\Fnuobs(\tobs)$
using the formulae of \S~3 and taking the integral~(\ref{eq:Ltot})
numerically. A simple illustrative afterglow model has the following 
parameters:

1. Isotropic energy $\Eej=10^{53}$~erg and initial Lorentz factor 
$\Gamma_0=200$ (the reverse shock is assumed to be non-relativistic
and $\Gamma_0\approx\Gej$).

2. External density $n_0(R)=const$ and $\mu_e=1$ (uniform hydrogen medium). 
We will consider $n_0=0.1,$ 1, 10, $10^2$, and $10^3$~cm$^{-3}$. 

3. Parameters of the postshock leptons: $\epe=0.1$ and $p=2.5$. 
Most of the blast-wave energy is carried by relativistically hot ions 
which do not exchange energy with the electrons and the blast wave is 
approximately adiabatic. 

4. Magnetic equipartition parameter $\epB$ immediately behind the shock.
It will be taken equal to $10^{-6}$, $10^{-4}$ or $10^{-2}$ in the example 
models. $\epB(m,\tR)$ can evolve in the postshock material of the blast 
wave as it expands and decelerates. We will consider two cases: magnetic 
flux is conserved in the postshock region (eq.~\ref{eq:tepB}) or 
$\epB(m,\tR)=\epB$ is constant in all shells $m$ at all times.

5. The $\gamma$-ray front is described by its isotropic energy $E_\gamma$.
In the examples below we assume $E_\gamma=\Eej=10^{53}$~erg, i.e., half of 
the total explosion energy is initially emitted in the prompt GRB. We also 
assume the standard GRB spectrum with a high-energy slope $\alpha_2=1.5$ 
which gives $\xiacc=120$ (\S~2).

The blast wave has a 
constant $\Gamma\approx\Gamma_0$ until it approaches the characteristic 
deceleration radius $\Rdec$ where the swept-up mass $m$ satisfies 
$\Gamma_0^2mc^2=\Eej$. At larger radii $\Gamma(R)$ quickly approaches the 
self-similar solution of Blandford \& Mckee (1976):
$\Gamma=(17/12)^{1/2}\Gamma_0(R/\Rdec)^{-3/2}$.
We will use a simple approximation: $\Gamma(R)=\Gamma_0$ at $R\leq\Rdec$ and 
$\Gamma(R)=\Gamma_0(R/\Rdec)^{-3/2}$ at $R\geq\Rdec$.
It has sufficient accuracy, adequate to our simplified hydrodynamical 
model (\S~3).

The results of numerical calculations are shown in Figures~3-7. 
In all examples, a cosmological redshift $z=1$ is assumed.  

\begin{figure}
\begin{center}
\plotone{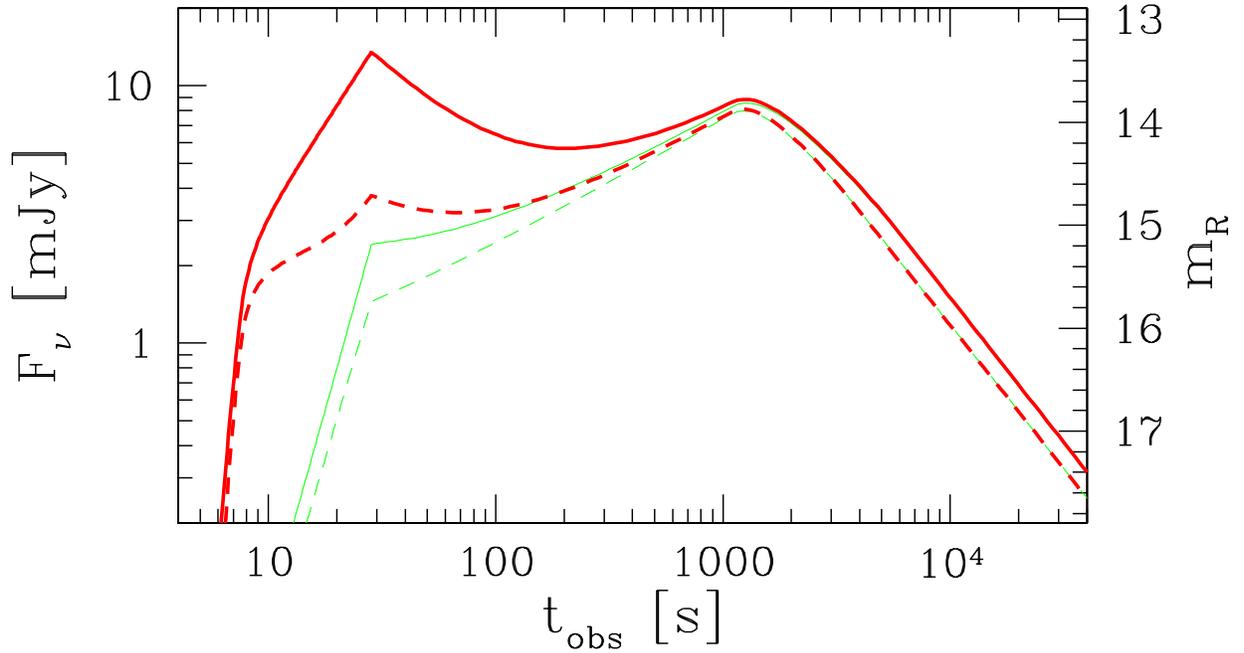}
\label{fig:f1}
\end{center}
\caption{ Example of the optical light curve in the R-band,
$\nuobs=5.45\times 10^{14}$~Hz, for a "canonical" GRB explosion:
adiabatic blast wave with 
$\Eej=E_\gamma=10^{53}$~erg, $\Gamma_0=200$, $z=1$, $n_0=10$~cm$^3$,
$\epB=10^{-4}$, $\epe=0.1$, $p=2.5$. 
Left axis indicates the observed flux in mJy and right axis --- the
corresponding R-magnitude.
Calculations shown by solid curves assume conservation of the 
postshock magnetic flux. Dashed curves show the results with
$\epB(m,R)=const=\epB$ assumption.
Thin light curves would be obtained if the
pairs were neglected ($Z=1$ and $\gamma=1$).
}
\end{figure}

\begin{figure}
\begin{center}
\plotone{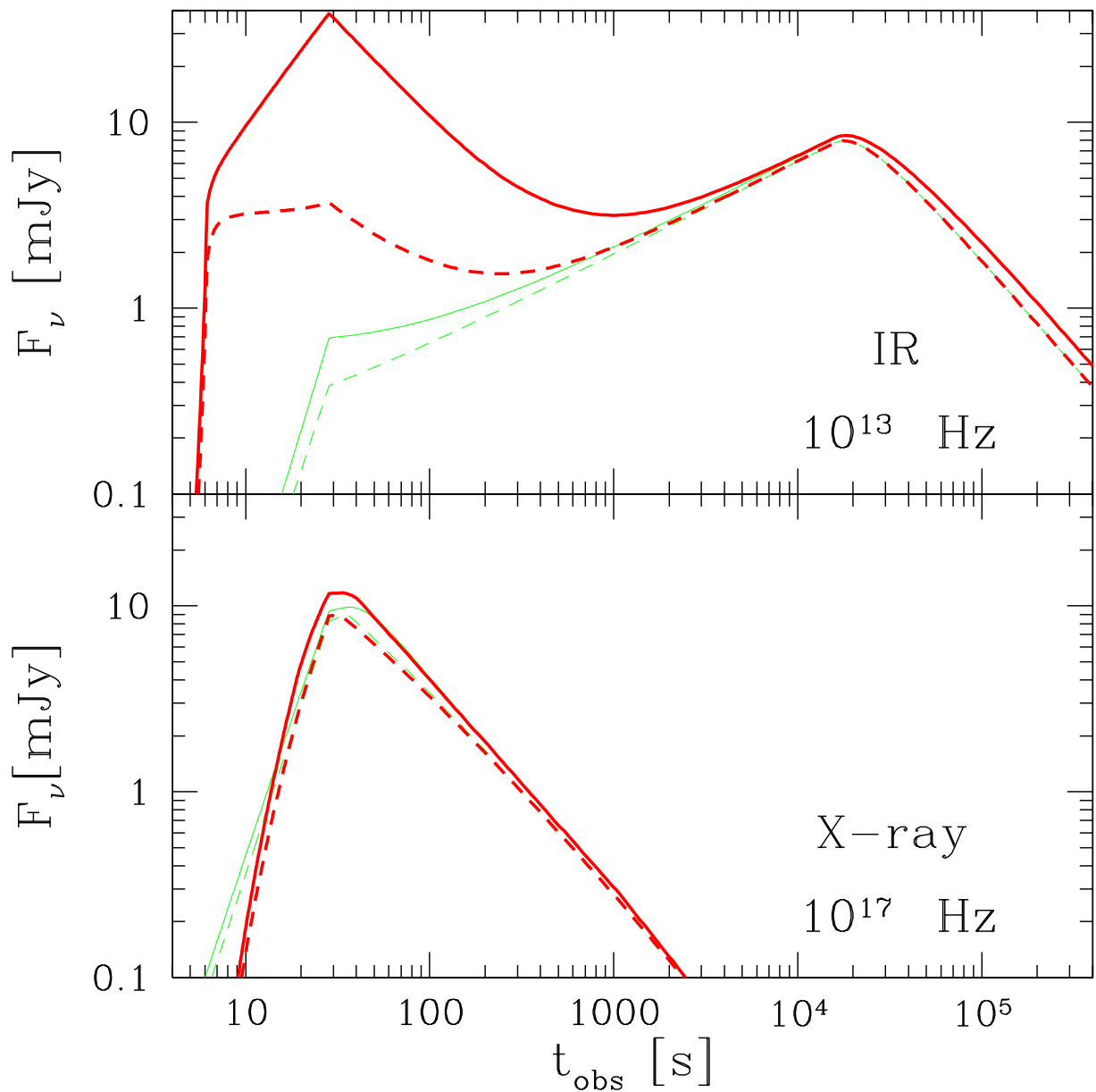}
\label{fig:f2}
\end{center}
\caption{Light curves at $\nuobs=10^{13}$~Hz (upper panel)
and $\nuobs=10^{17}$~Hz (lower panel) for the same model as in 
Figure~3. The meaning of line types is the same as in Figure~3.
}
\end{figure}

\begin{figure}
\begin{center}
\plotone{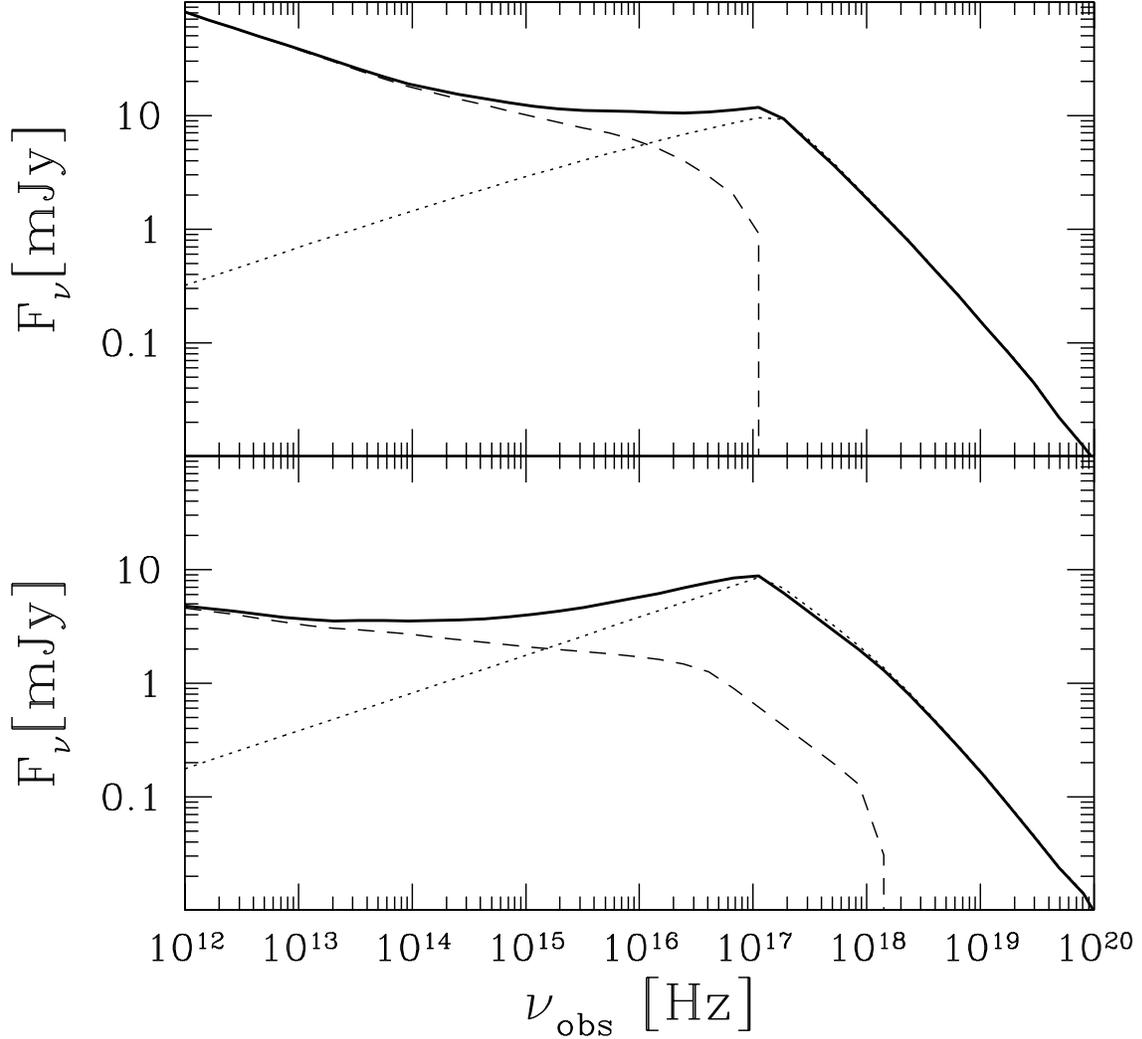}
\end{center}
\caption{Instantaneous spectrum emitted by the blast wave at $R=\Rdec$
is shown by the solid curve (the model parameters are the same as in 
Fig.~3). Dashed curve shows the contribution of the 
$e^\pm$-loaded shell $m<\mload$. Dotted curve shows the spectrum that 
would be found in the absence of $\gamma$-ray front ($Z=1$ and 
$\gamma=1$); it is approximately equal to the spectrum of pair-free 
material $m>\mload$.
Upper panel: conservation of postshock magnetic flux is assumed.
Lower panel: $\epB(m,\tR)=const=\epB$ is assumed.
}
\end{figure}

\begin{figure}
\begin{center}
\plotone{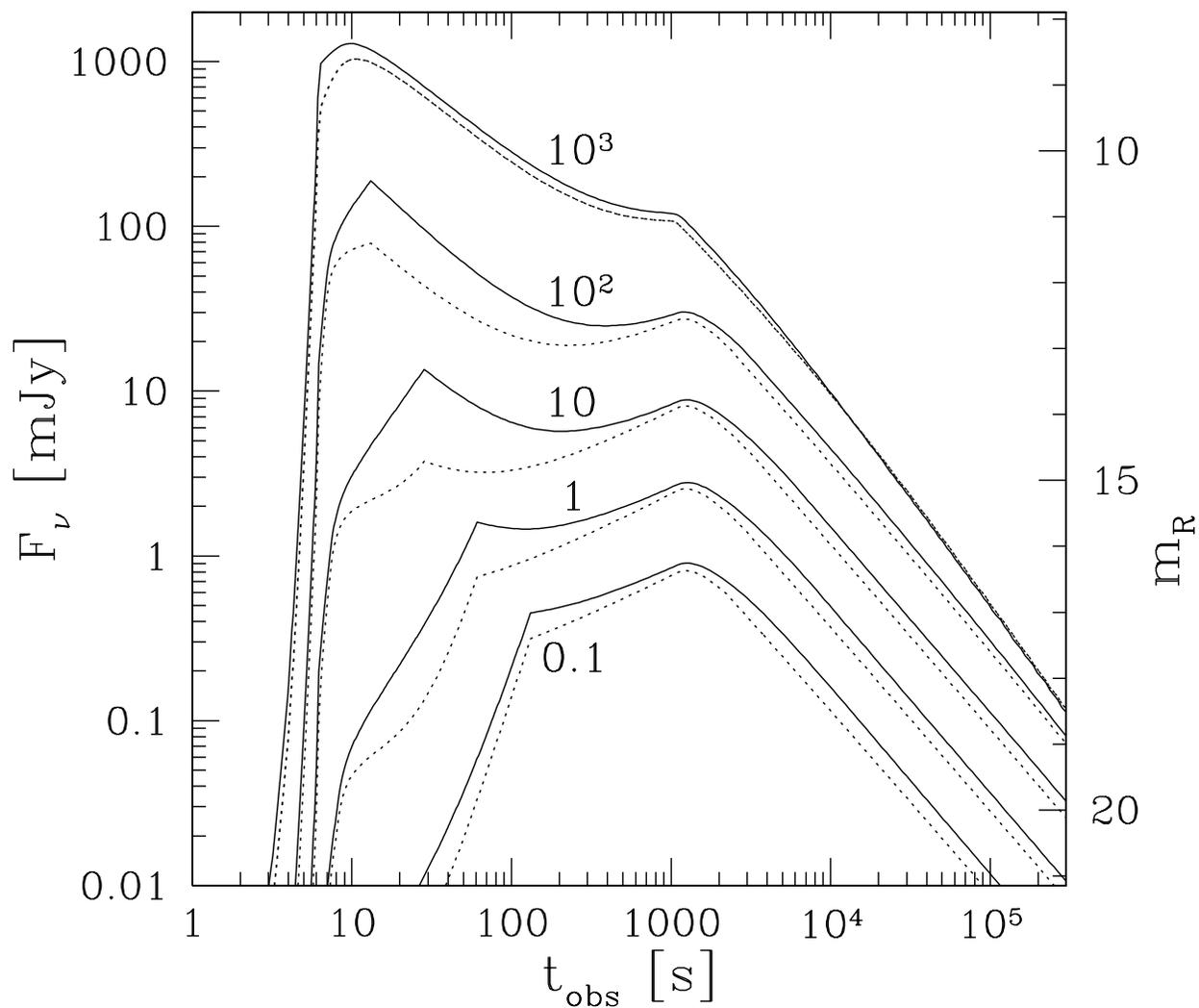}
\label{fig:f8}
\end{center}
\caption{ Optical light curves found for external densities $n_0=0.1$, 
1, 10, $10^2$, and $10^3$~cm$^{-3}$. All other parameters are the same as 
in Figure~3. Calculations shown by solid curves assume conservation of the 
postshock magnetic flux. Dotted curves show the results with
$\epB(m,R)=const=\epB$.
}
\end{figure}

\begin{figure}
\begin{center}
\plotone{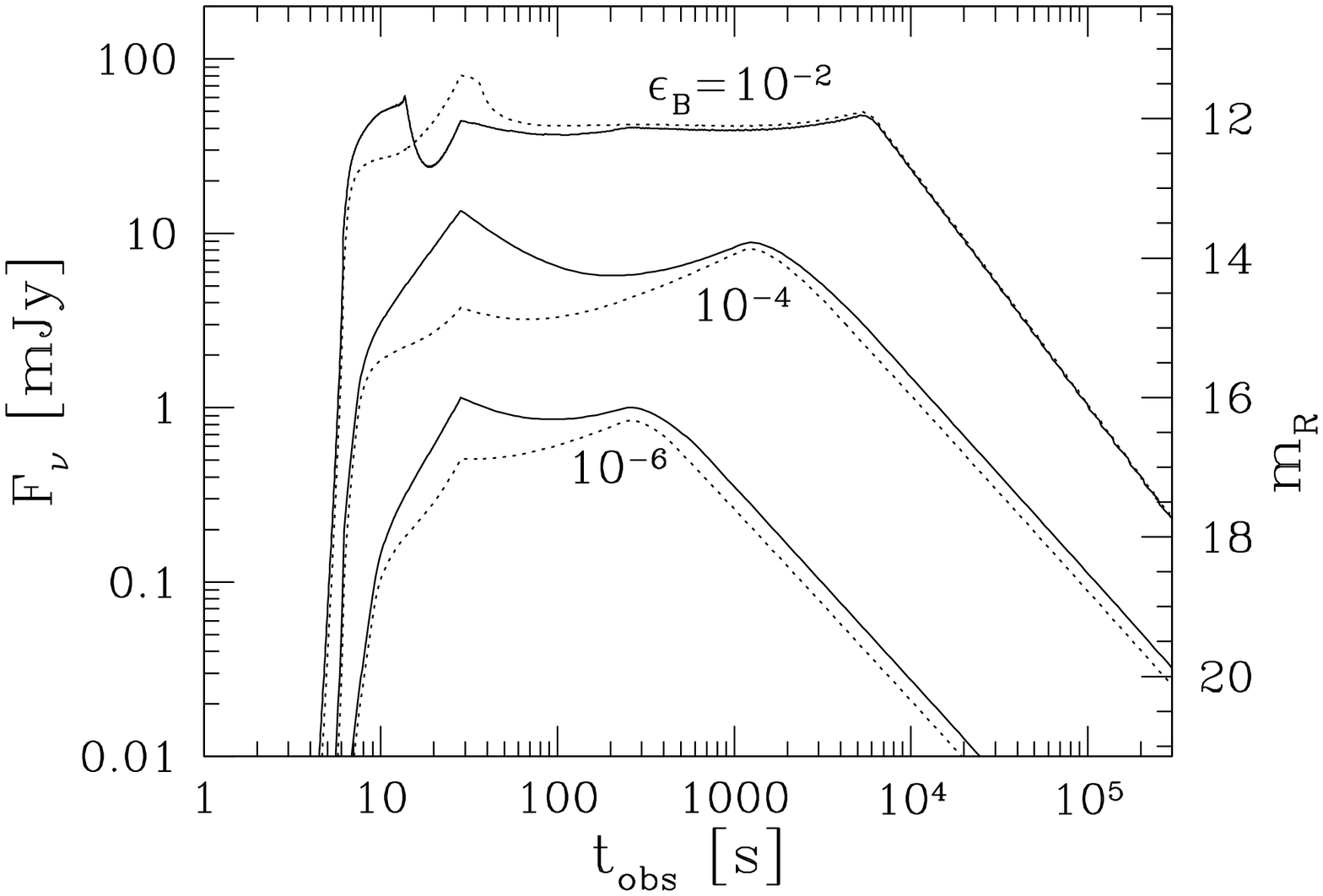}
\label{fig:f9}
\end{center}
\caption{ Optical light curves found for 
$\epB=10^{-6}$, $10^{-4}$, and $10^{-2}$. 
All other parameters are the same as in Figure~3.
Calculations shown by solid curves assume conservation of the postshock
magnetic flux. Dotted curves show the results with $\epB(m,R)=const=\epB$.
}
\end{figure}

Figure~3 shows the optical light curve $\Fnuobs(\tobs)$ found at 
$n_0=10$~cm$^{-3}$ and $\epB=10^{-4}$. It has two peaks. The first peak 
is a result of $e^\pm$ loading 
(for comparison we show the results with $Z=1$ and $\gamma=1$, i.e.,
with neglected impact of the $\gamma$-ray front on the external medium).
The figure compares the results obtained with magnetic flux conservation 
and $\epB(m,\tR)=const$. The $e^\pm$ emission is stronger if the 
magnetic flux is conserved because it implies a higher $\tepB$; then the
peak of $e^\pm$ radiation is reached at $\tR=\Rdec$ (see \S~5). 

The second peak is produced by the pair-free part of the blast.
It is well described by the standard afterglow model and 
corresponds to $\nu_m=\nu$ (see \S~5.1). The time of the second
peak $t_{\rm peak}\propto (\epB n_0)^{1/3}\epe^{4/3}\nu_{15}^{-2/3}$.

Figure~4 shows the corresponding light curves in the infrared and X-ray 
bands. The relative contribution of $e^\pm$ 
to the afterglow is higher at lower frequencies, and their effect is 
negligible in the hard X-ray band. Comparing the light curves in IR, 
optical, and soft X-rays, one can see that the spectral flux of the $e^\pm$ 
peak is comparable in all three cases, indicating a small spectral index of 
$e^\pm$ radiation. 

Figure~5 shows the instantaneous broad-band spectrum emitted by the blast 
wave at the deceleration radius $\Rdec$ (which is 4.6 larger than $\Racc$
in the example model). 
The small $e^\pm$-dominated fraction of the swept-up mass dominates the 
emission in soft bands. Its spectral slope is small: $|\alpha|\simlt 0.2$.
At high frequencies, $\nuobs>10^{17}$~Hz, the spectrum is dominated by 
the pair-free material behind the current position of the shock
and well described by the usual pair-free model.
In particular, the standard breaks at $\nu_m\sim 10^{18}$~Hz 
[$\nu^{1/3}\rightarrow\nu^{-(p-1)/2}$] and $\nu_c\sim 10^{20}$~Hz 
[$\nu^{-(p-1)/2}\rightarrow\nu^{-p/2}$] are seen.
[In the upper panel, the second (cooling) break occurs at a smaller 
frequency $\sim 10^{19}$~Hz and is smooth because $\nu_c$ happens to be 
close to $\nu_m$.] The spectrum of the $e^\pm$ emission component has a 
break at $\nu\sim\nu_m(\mload,\tR)$, which is $\sim 10^{18}$~Hz in the 
upper panel and $\sim 3\times 10^{17}$~Hz in the lower panel, and cuts 
off at $\nu>\nu_c(\mload,\tR)$. The cutoff frequency is smaller in the 
upper panel because the flux conservation gives a stronger magnetic field.

Figure~6 shows the optical light curves for different $n_0$ and fixed 
$\epB=10^{-4}$. It also illustrates the effect of magnetic flux conservation 
by comparing it with the results obtained at $\epB(m,\tR)=const$. The $e^\pm$ 
emission component quickly increases with increasing $n_0$. It scales as 
$n_0^{3/2}$ as we show in \S~5, and this scaling is observed in Figure~6 for 
constant-$\epB$ models.

Figure~7 shows how the optical light curve depends on the shock parameter 
$\epB$. At $\epB=10^{-2}$ the $e^\pm$ are in the fast-cooling regime 
($\nu_c<\nu_m$) and most of the pair energy is emitted at earlier times. 
This produces a bump in the light curve before or at the deceleration
radius, depending on the behavior of magnetic field downstream of the 
shock.


\section{Analytical calculation}

In \S~5.1 we derive the luminosity of a relativistic blast wave without 
a $\gamma$-ray precursor, when the $e^\pm$-loading and preacceleration 
effects are neglected ($Z=1$ and $\gamma=1$). The pair-free blast wave was
studied in a number of previous works (e.g. Sari, Piran, \& Narayan 1998;
Granot, Piran, \& Sari 1999; Panaitescu \& Kumar 2000) and used extensively
to fit observed afterglows. We show that the previous results can be obtained
with a different technique where the produced luminosity is calculated as 
an integral over the Lagrangian mass coordinate $m$ (eq.~\ref{eq:Ltot}).
This approach is extented to $e^\pm$-loaded blast waves in \S~5.2. 

Where the ambient density profile needs to be specified we assume a power-law,
so that the swept-up mass is
\be
\label{eq:k}
  m(R)=\macc\left(\frac{R}{\Racc}\right)^k.
\ee
$k=3$ describes a uniform medium $n_0(R)=const$ and $k=1$ describes a 
wind-type medium $n_0\propto R^{-2}$. The derivation in \S\S~5.1 and 5.2 
applies to both cases. In \S~6 we consider in more detail the case $k=3$.

\subsection{Pair-free afterglow}

From \S~3.4 we have at $Z=1$ and $\gamma=1$,
\begin{eqnarray}
\label{eq:dL0}
  L_\nu^0=\int_0^{\tm} 
  30\,\left(\frac{\tepB\tn_0}{\mu_e}\right)^{1/2}
  \left\{\begin{array}{ll}
    (\nu/\tnu_m)^{1/3} & \nu<\tnu_m<\tnu_c \\
    (\nu/\tnu_m)^{(1-p)/2} & \tnu_m<\nu<\tnu_c \\
    (\nu/\tnu_c)^{1/3} & \nu<\tnu_c<\tnu_m \\
    0 & \nu>\tnu_c
  \end{array}\right\}\;\dd m.
\end{eqnarray}
Symbol "0" in $L_\nu^0$ marks the neglect of $e^\pm$ loading.
Three quantities in the integrand depend on the Lagrangian coordinate $m$:
$\tepB$, $\tg_m$, and $\tg_c$. Quantities $\tepB(m)$ and $\tnu_m(m)$ can 
vary significantly in the region $m\ll\tm$, however, this region makes a 
negligible contribution to the integral. Therefore, one can approximate 
$\tepB$ and $\tnu_m$ as constants: $\tepB(m)\approx \tepB(\tm)$ and 
$\tnu_m(m)\approx\tnu_m(\tm)$. Only $\tnu_c(m)$ cannot be assumed constant: 
$\tnu_c(m)\rightarrow\infty$ at $m\rightarrow\tm$. 

A characteristic $\tnu_c$ can be defined at $m=\tm/2$: it will represent 
the cooling cutoff in the main part of the postshock material. The value 
of $\tg_c(\tm/2)$ is shaped as the blast wave expands from
$R=2^{-1/k}\tR$ (the shock radius of $\tm/2$) to the current radius $\tR$. 
It is given by (see \S~3.3),
\be
\label{eq:gc_}
  \tg_c(\tm/2)=\frac{a\; m_e}{\sT m_p\,\mu_e\, n_0\epB \tG\tR },
\ee
where $a\approx (3/16)(1-2^{-1/k})^{-1}$ is a numerical factor.
We assume here a small Compton factor $C<1$ for simplicity.

Let us define 
$\gamma_\nu\equiv\tg_\nu(\tm/2)\approx\tg_\nu(\tm)$,
$\gamma_m\equiv\tg_m(\tm/2)\approx\tg_m(\tm)$, and 
$\gamma_c\equiv\tg_c(\tm/2)$;
they are given by equations~(\ref{eq:tgnu}), (\ref{eq:tgm}), and 
(\ref{eq:gc_}), respectively. The numerical values of these Lorentz factors 
and the corresponding synchrotron frequencies are\footnote{
Hereafter in \S~5.1 we drop tilde for $\tR$, $\tepB$, and $\tG$ 
because there is no need to consider $m\ll \tm$ for the pair-free afterglow
and distinguish between the postshock and current quantities.}
\beq
\label{eq:gnu}
  \gamma_\nu &=& 2.7\times 10^2\,\nu_{15}^{1/2}\,\Gamma_2^{-1}\,
  (\epB n_0\mu_e)^{-1/4},  \\
\label{eq:gc}
  \gamma_c   &=& 8.2\times 10^2\, a\, R_{16}^{-1}\,\Gamma_2^{-1}\,
   (\epB n_0\mu_e)^{-1},  \qquad
   \nu_c =9.2\times 10^{15}\,a^2\left(\epB n_0\mu_e\right)^{-3/2}
          R_{16}^{-2} {\rm ~Hz}, \\
\label{eq:gm}
  \gamma_m   &=& 1.84\times 10^5\,\Gamma_2\,\psi\,\mu_e,  \qquad
 \hspace*{2.2cm} \nu_m = 4.6\times 10^{20} \,\Gamma_2^4
     \left(\epsilon_B\,n_0\right)^{1/2}\psi^2\mu_e^{5/2} {\rm ~Hz},
\eeq
where 
\be
\label{eq:psi}
 \psi\equiv\frac{\epe(p-2)}{(p-1)}.
\ee

There are two possible cases:

{\bf Case I}. $\nu_c>\nu$. Most of the blast material contributes to the 
observed luminosity at the frequency $\nu$. The luminosity integral over 
$m$ is then estimated as $\nu L_\nu\approx \tm(\delta L/\delta m)$ where 
$\delta L/\delta m$ is taken at $m=\tm/2$ and given by 
equation~(\ref{eq:dL0}) with $\tnu_c$ and $\tnu_m$ replaced 
by $\nu_c$ and $\nu_m$.

{\bf Case II}. $\nu_c<\nu$. Most of the blast material does not emit any 
synchrotron radiation at $\nu$ because its spectrum is cut off at a lower 
frequency $\nu_c$. Only a mass fraction $(\tm-m_c)\ll \tm$ just behind 
the shock front contributes to $L_\nu$, where $m_c$ is defined by 
$\tnu_c(m_c)=\nu$. The condition $\tnu_c>\nu$ is satisfied in the mass 
interval $m_c<m<\tm$ and the luminosity integral over $m$ can be estimated 
as $\nu L_\nu\approx (\tm-m_c)(\delta L/\delta m)$.  

The cooling Lorentz factor $\tg_c(m)$ increases toward the shock front 
$\tm$ as $\tg_c(m)\propto(\tR-R)^{-1}\propto (\tm^{1/k}-m^{1/k})^{-1}$, 
and we have 
\be
\label{eq:b}
  \frac{\tg_c(\tm/2)}{\tg_c(m)}=\frac{1-(m/\tm)^{1/k}}{1-(1/2)^{1/k}}
     =b\,\left(1-\frac{m}{\tm}\right)
\ee
where so defined $b(m)$ is a slowly varying function: 
$b=2$ at $m=\tm/2$ and $b=[k(1-0.5^{1/k})]^{-1}$ at $m\rightarrow\tm$. 
Equating $\tg_c(m)$ and $\gamma_\nu$, we find $m_c$, 
\be
   \frac{\tm-m_c}{\tm}=\frac{\gamma_c}{b\gamma_\nu}=
   \frac{1}{b}\left(\frac{\nu_c}{\nu}\right)^{1/2}, 
\ee
with $b=2$ at $\nu_c=\nu$ and $b=[k(1-0.5^{1/k})]^{-1}$ at 
$\nu_c\ll\nu$. One can take $b\approx 2$ in all cases.

\medskip

The results can be summarized as follows,
\begin{eqnarray}
\label{eq:L0}
 L_\nu^0 &\approx & 30\,\left(\frac{\epB n_0}{\mu_e}\right)^{1/2}
                  \,\tm\;g_\nu,\\
\label{eq:g}
  g_\nu &=& \frac{1}{1+b(\nu/\nu_c)^{1/2}} \times 
   \left\{\begin{array}{ll}
    (\nu/\nu_m)^{1/3}     & \nu<\nu_m<\nu_c, \\
    (\nu/\nu_c)^{1/3}     & \nu<\nu_c<\nu_m, \\
    (\nu/\nu_m)^{(1-p)/2} & \nu_m<\nu<\nu_c, \\
    1                     & \nu_c<\nu<\nu_m, \\
    (\nu/\nu_m)^{(1-p)/2} & \nu_m<\nu_c<\nu, \\
    (\nu/\nu_m)^{(1-p)/2} & \nu_c<\nu_m<\nu.
  \end{array}\right.
\end{eqnarray}
It agrees with the usual afterglow description (e.g. Sari et al. 1998).

After the deceleration radius, 
we have for an adiabatic blast wave $\Gamma^2 m c^2=\Eej$ and then 
\be
  \Fnuobs(R)=\frac{\Eej(1+z)}{2\pi D^2 c^2}\frac{\dLmax_\nu}{\delta m}\; g_\nu
  \approx 0.3\,(1+z)\,\left(\frac{\Eej}{10^{53}\rm erg}\right)
               \left(\frac{D}{10^{28}\rm cm}\right)^{-2}
         \left(\frac{\epsilon_B n_0}{\mu_e}\right)^{1/2}\; g_\nu
   \; {\rm Jy}.
\ee
One can see from this equation that 
the observed flux at a fixed $\nu$ scales as $n_0^{1/2}\,g_\nu$.
It can be found as a function of observer time $\tobs$
using the approximate relation $\tobs\approx (R/2\Gamma^2 c)(1+z)$ and 
the deceleration law for an adiabatic blast wave, $\Gamma^2=\Eej/mc^2$,
which gives
\be
   \tobs\approx \frac{R mc}{2\Eej}\,(1+z).
\ee

\subsection{Pair-loaded afterglow}

We now calculate the instantaneous luminosity of the blast wave $L_\nu(\tR)$ 
taking into account $e^\pm$ loading.
The pairs dominate the material with Lagrangian coordinate $m<\mload$.
The luminosity from this material can be written as
\begin{eqnarray}
\label{eq:Lpm_d}
 L_\nu^\pm(\tR) & = &\int_0^{m_1} \frac{\delta L_\nu}{\delta m}\,\delta m,
                \qquad m_1=\min\{\mload,\tm\}.
\end{eqnarray}
At radii $\tR<\Rload$ ($\tm<\mload$) the whole blast
is $e^\pm$ dominated and its total luminosity $L_\nu=L_\nu^\pm$. 
At radii $\tR>\Rload$ ($\tm>\mload$) the total luminosity is a sum of two 
parts: $L_\nu^\pm$ from small $m<\mload$ and luminosity $L_\nu^0$ from 
the $e^\pm$-free material $\mload<m<\tm$. The latter peaks near $\tm$
(\S~5.1).

The total luminosity can be written as the sum,
\be
\label{eq:sum}
   L_\nu=L_\nu^\pm +L_\nu^0.
\ee
At small radii $\tR<\Rload$ we have just one integral $L_\nu=L_\nu^\pm$,
however, $L_\nu^0$ may be formally kept in equation~(\ref{eq:sum}) and
interpreted as a luminosity that would be obtained at $Z=\gamma=1$.
It would not change $L_\nu$ because $L_\nu^0\ll L_\nu^\pm$ at $\tR<\Rload$ 
in the soft bands of interest.

\subsubsection{Calculation of $L_\nu^\pm$}

\begin{figure}
\begin{center}
\plotone{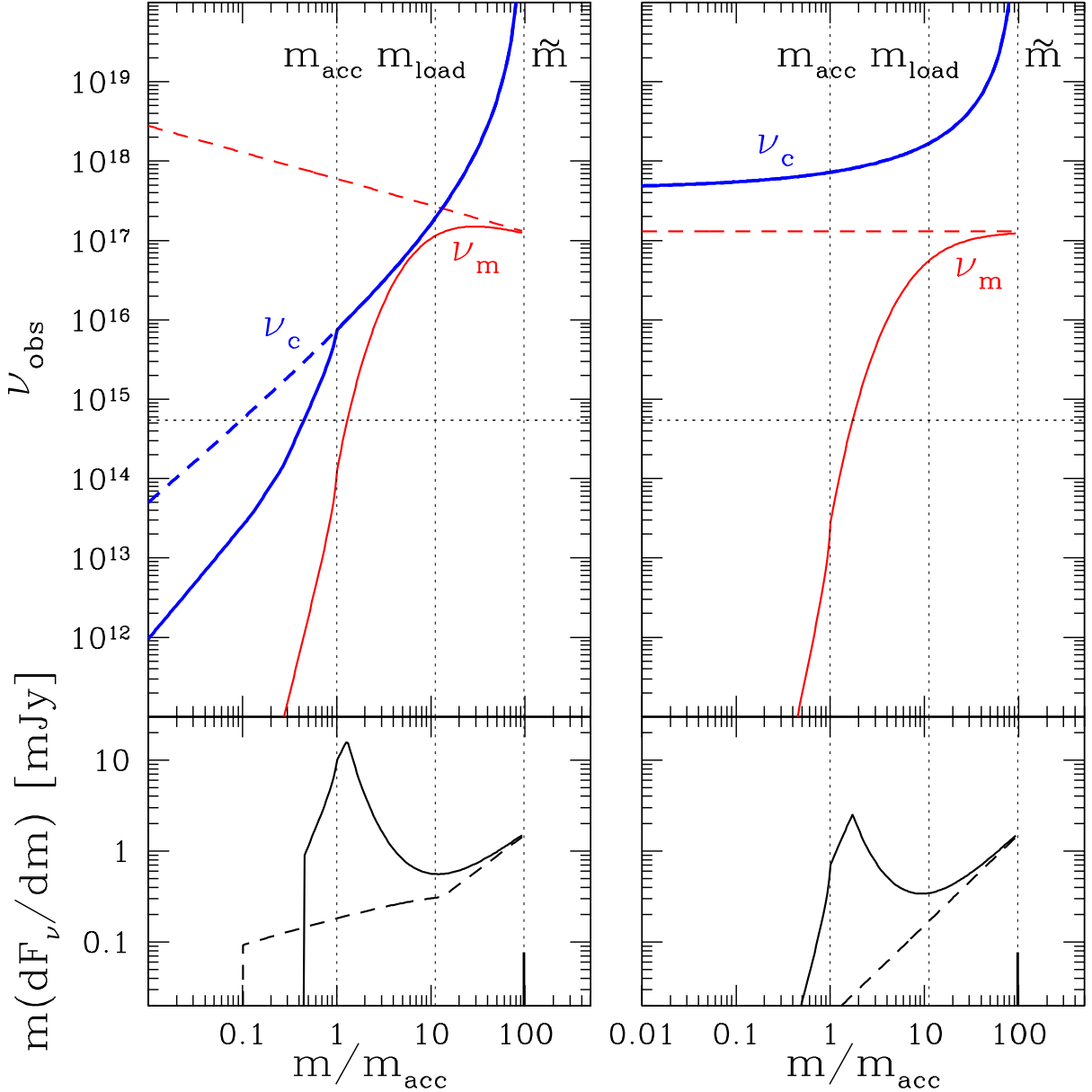}
\end{center}
\caption{\small 
Instantaneous blast-wave emission as a function of Lagrangian 
coordinate $m$ at radius $\tR=\Rdec$. The blast-wave model is the same 
as in Figure~3. 
The results that would be obtained without the $\gamma$-ray 
front effect ($Z=1$ and $\gamma=1$) are shown by dashed curves.
Upper panel: Observed synchrotron peak frequency $\nu_m(m,\Rdec)$
and cutoff frequency $\nu_c(m,\Rdec)$ corrected for cosmological 
redshift (i.e. divided by $1+z$). The horizontal dotted 
line indicates the $R$-band frequency $\nuobs=5.45\times 10^{14}$~Hz. 
Lower panel: Distribution of the produced optical flux over $m$. 
The emission from the $e^\pm$-loaded mass $m<\mload$ sharply peaks near 
$\macc$, at a specific $m_*$ where $\tnu_m/(1+z)=\nuobs$. The pair-free 
emission component peaks at the current position of the shock $m=\tm=\mdec$.
{\it Left:} Conservation of postshock magnetic flux is assumed. 
{\it Right:} $\epB(m,\tR)=const=\epB$ is assumed.
}
\end{figure}

Pair-loaded blast waves have a special feature: the peak synchrotron 
frequency $\nu_m(m,\tR)$ varies enormously with the Lagrangian coordinate 
$m$ at $m<\mload$ (see an example in Figure~8). The peak frequency is given
by
\be
\label{eq:tnu_m}
  \tnu_m(m)\equiv\nu_m(m,\tR)=0.2\tG\,\frac{e\tB}{m_ec}\,\tg_m^2
   =4.6\times 10^{12}\,\frac{\tG^3\Gamma\,\tn_0\,\tepB^{1/2}\psi^2\mu_e^{5/2}}
    {\tg(1+\tb)[\gamma(1+\beta)]^{3/2}Z^2\,n_0^{1/2}} {\rm ~Hz},
\ee
where we have taken into account the adiabatic cooling factor $A$ (\S~3.2) 
and assumed a slow radiative cooling of $e^\pm$. The factor 
$\gamma^{-3/2}Z^{-2}$ appearing in this expression is a very steep function 
of $m$, $R(m)$, or $\xi(R)$, whichever is chosen as a Lagrangian coordinate 
in the blast wave. It varies by five orders of magnitude when $\xi$ varies 
from $\xiacc/2$ to $2\xiacc$, which corresponds to a narrow range of $R$ 
from $\Racc/\sqrt{2}$ to $\sqrt{2}\Racc$.

The steep variation of $\tnu_m$ with $m$ implies that a narrow mass
shell $\Delta m\sim\macc$ 
dominates $L_\nu^\pm$ in a broad range of $\nu$. This fact enables a simple 
estimate of $L_\nu^\pm$. To a first approximation, the number of $e^\pm$ 
whose emission peaks at a given $\nu$ does not depend on $\nu$ and equals
the number of particles in the shell $\Delta m\sim \macc$,
\be
\label{eq:N_nu}
 N_*\sim \Zacc\frac{\macc}{\mu_em_p}=74\,\frac{\macc}{m_p}.
\ee
The produced luminosity
$\dd L^\pm/\dd\log\nu$ approximately equals 
$\dot{E}N_*/2$, which gives the spectral luminosity,
\beq
 L^\pm_\nu(\tR) &=& \frac{5}{12\pi}\frac{m_ec^2\sT}{e}
   \frac{\tB_{\rm acc}}{\tG}N_*
  =30\left(\frac{\tep_{B\rm acc}\tn_0}{\mu_e}\right)^{1/2}\Zacc\macc\,f_\nu, 
  \qquad \tR>\Racc. 
\label{eq:Lpm}
\eeq
Here $\tB_{\rm acc}$ is the current magnetic field in the shell $\Delta m$ 
and $\tep_{B\rm acc}$ is the corresponding magnetic parameter
($\tB_{\rm acc}^2/8\pi$ divided by the current energy density of the blast 
wave). In the first approximation, $f_\nu=1$, i.e. $L^\pm_\nu$ does not 
depend on $\nu$ and the spectral index of emission $\alpha\approx 0$. 

Calculation of $L^\pm_\nu$ by accurate integration over $m$ gives the 
correction factor $f_\nu$ and confirms that $f_\nu\sim 1$ in a broad 
range of frequencies. The correction factor is derived analytically 
in Appendix and shown in Figure~9. 
It is conveniently expressed as a function of
$\nu/\tnuacc$ where $\tnuacc\equiv\nuacc(\tR)\equiv\nu_m(\macc,\tR)$
(thus, $\tnuacc$ describes the peak 
of synchrotron emission from material shocked at $R=\Racc$ after it 
expanded to the current $\tR$),
\be
\label{eq:nuacc}
  \tnuacc=4\times 10^{14}\,\tG_2^2
    \left[\frac{\tG}{\Gamma(\Racc)}\right]
    \left[\frac{\tn_0}{n_0(\Racc)}\right]^{1/2}
    \left(\frac{\tn_0\,\tep_{B\rm acc}}{0.01}\right)^{1/2}
    \left(\frac{\psi}{0.1}\right)^2 {\rm ~Hz}, \quad \tR\geq\Racc.
\ee
This frequency is near the optical band. 

\begin{figure}
\begin{center}
\plotone{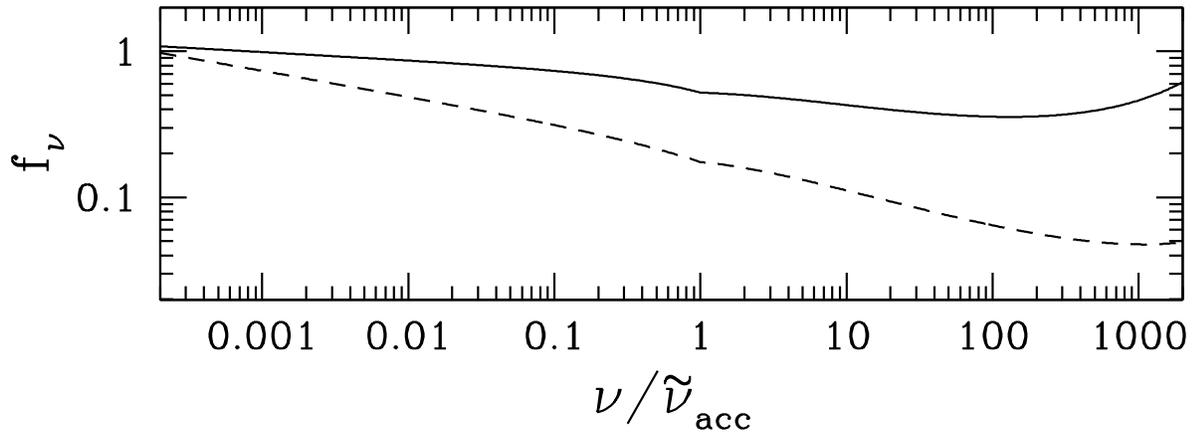}
\end{center}
\caption{ Correction factor $f_\nu$ (with $\tep_{B*}=\tep_{B\rm acc}$,
see Appendix). Solid curve shows the case of $k=3$ (uniform medium),
and dashed curve --- $k=1$ (wind-type medium). The characteristic
frequency $\tnuacc$ is given by equation~(\ref{eq:nuacc}).
}
\end{figure}

The basic reason for the small slope $\alpha$ is the steep dependence of the 
pair-loading process on radius (the big power 17/2 appears in the problem).

\subsubsection{$e^\pm$ contribution to the observed luminosity}

The total observed flux of the blast-wave radiation is (see eqs.~\ref{eq:F} 
and \ref{eq:sum})
\be
  \Fnuobs=\Fnuobs^\pm+\Fnuobs^0
         =\frac{(1+z)\tG^2}{3\pi D^2}\left(L_\nu^\pm+L_\nu^0\right),
\ee
where the pair-dominated part is given by equation~(\ref{eq:Lpm}) 
and the pair-free part by equation~(\ref{eq:L0}). This analytical result 
is compared with the numerical calculations in Figure~10. The analytical 
calculation of $\Fnuobs^\pm$ is more accurate because the integral peaks 
in a narrow and well defined region of $m$. The standard pair-free part
is still reasonably well approximated and we can compare the two components 
analytically.

\begin{figure}
\begin{center}
\plotone{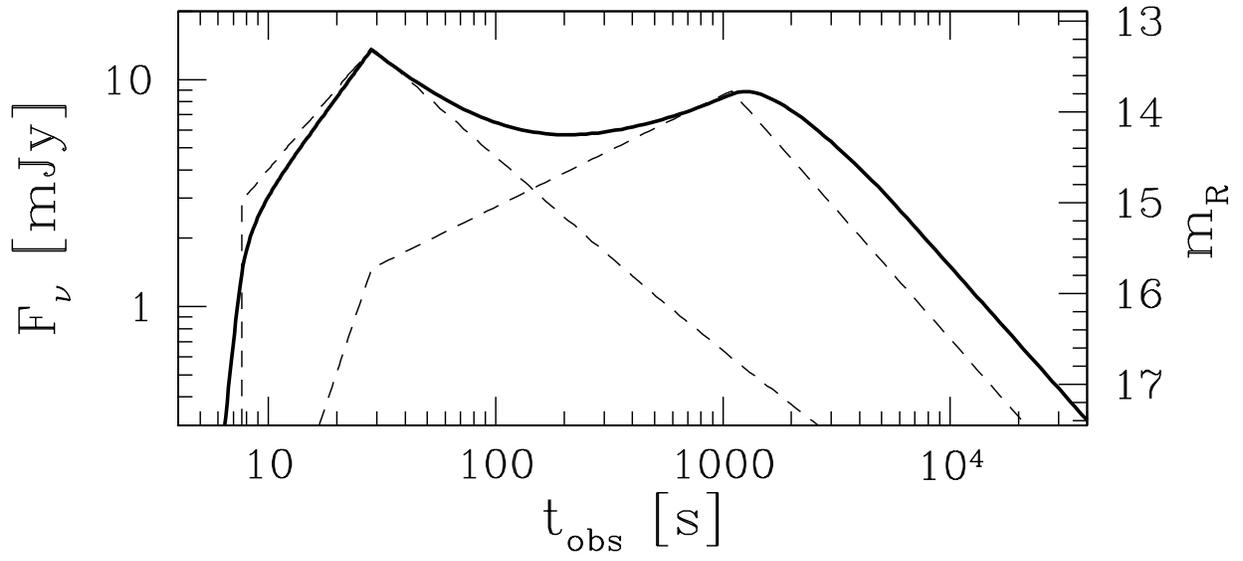}
\end{center}
\caption{ Optical light curve calculated numerically in Figure~3 is
here compared with the two components $\Fnuobs^\pm$ and $\Fnuobs^0$
calculated analytically (dashed lines). 
}
\end{figure}

Using equations~(\ref{eq:L0}) and (\ref{eq:Lpm}), we get
\be
\label{eq:ratio}
    \frac{\Fnuobs^\pm}{\Fnuobs^0}(\tR)=74\mu_e\,\frac{\macc}{\tm}
   \,\frac{f_\nu}{g_\nu}\,\left(\frac{\tep_{B\rm acc}}{\epB}\right)^{1/2},
  \qquad  \tR\geq\Racc.
\ee
$\Fnuobs^0$ peaks in the X-ray band during the early afterglow, and 
$g_\nu\ll 1$ in IR, optical, and UV bands. 
By contrast, $\Fnuobs^\pm$ has the spectral factor $f_\nu\sim 1$
in a broad range of soft frequencies from IR to soft X-rays (\S~5.2.2).

The $e^\pm$-loaded material dominates the observed flux in the soft bands 
for two reasons: (1) the number of emitting particles is increased by the 
factor $Z\sim \Zacc=74\mu_e$ and (2) their synchrotron peak frequency is 
shifted to soft bands (because the mean energy per postshock particle is 
reduced). In addition, $\tep_{B\rm acc}>\epB$ if the postshock magnetic 
flux is conserved in the blast wave (see \S~3.2). As a result, 
$\Fnuobs^\pm\gg \Fnuobs^0$ even when the blast wave expands to $\tR>\Rload$ 
where the $e^\pm$-loaded fraction of the blast material is very small, 
$\mload\ll\tm$.


\section{Blast wave in a uniform medium}

We now consider in detail the case of a uniform hydrogen medium:
$n_0(R)=const$ ($k=3$) with $\mu_e=1$, and aim to find the observed flux 
$\Fnuobs=\Fnuobs^\pm+\Fnuobs^0$ as a function of $\tobs$ and $\nuobs$.
We shall assume $\Rload<\Rdec$ where
\be
\label{eq:Rdec}
  \Rdec=\left(\frac{3\Eej}{4\pi\Gamma_0^2n_0m_pc^2}\right)^{1/3}
              =1.17\times 10^{17}\,
               \left(\frac{\Eej}{10^{53}\rm erg}\right)^{1/3}
               \left(\frac{\Gamma_0}{100}\right)^{-2/3}n_0^{-1/3}\;{\rm cm}.
\ee
We shall also assume that the blast wave is approximately adiabatic
and its Lorentz factor is given by 
\be
 \tG=\Gamma_0\times\left\{\begin{array}{ll}
    1 & \tR < \Rdec, \\
  (\tR/\Rdec)^{-3/2} & \tR > \Rdec. \\
                    \end{array}
             \right.
\ee
The approximate relation between the observer time $\tobs$,
the blast wave radius $\tR$, and its Lorentz factor $\tG$ reads (see \S~4),
\be
  \tR(\tobs)\approx 2\tG^2\,c\,\frac{\tobs}{1+z}=\left\{\begin{array}{ll}
    2\Gamma_0^2c(\tobs/1+z)  & \tobs < \tdec, \\
    \Rdec(\tobs/\tdec)^{1/4} & \tobs > \tdec, \\
                                         \end{array}
                                  \right.
\ee
where 
\beq
\label{eq:tdec}
  \tdec = (1+z) \frac{\Rdec}{2\Gamma_0^2c}
 \approx 194\,(1+z)\left(\frac{\Eej}{10^{53}{\rm ~erg}}\right)^{1/3}
       \left(\frac{\Gamma_0}{100}\right)^{-8/3}\,n_0^{-1/3}\;{\rm s},
\eeq
is the deceleration time.
We will use below the following explicit relations at $\tR>\Rdec$,
\be
   \tR(\tobs)=3.1\times 10^{16}\,
    \left(\frac{\Eej}{10^{53}{\rm ~erg}}\right)^{1/4}\,n_0^{-1/4}
    \left(\frac{\tobs}{1+z}\right)^{1/4} \; {\rm cm},
\ee
\be
\label{eq:G}
   \tG(\tobs)=\Gamma_0\left(\frac{\tobs}{\tdec}\right)^{-3/8}
             =720\, \left(\frac{\Eej}{10^{53}{\rm ~erg}}\right)^{1/8}
              \,n_0^{-1/8} \left(\frac{\tobs}{1+z}\right)^{-3/8}.
\ee

If the postshock magnetic flux is conserved in the expanding blast wave, 
the evolution of $\epsilon_B$ is described by equation~(\ref{eq:tepB}),
\be
\label{eq:tepB_homog} 
    \epsilon_B(m,\tR)
     =\epsilon_B\left(\frac{\tG^2}{\Gamma_0\Grel}\right)^{1/2}
                  \left(\frac{\tR}{R}\right)^2
     =\epsilon_B\left[\gamma(1+\beta)\right]^{1/2}
\left\{\begin{array}{ll}
    (\tobs/\tdec)^2       & \tobs<\tdec, \\
    (\tobs/\tdec)^{1/8} & \tobs>\tdec,
                                \end{array}
                         \right.
\ee
where $\epsilon_B$ is the magnetic parameter immediately behind 
the shock when the blast wave had radius $R(m)$, and $\beta(m)$ is the 
preshock medium velocity at that moment.

\subsection{Pair-free part}

The $e^\pm$-free afterglow is described by equations~(\ref{eq:F})
and (\ref{eq:L0}),
\be
\label{eq:Fnu0}
  \Fnuobs^0=\frac{\tG^2(1+z)}{3\pi D^2}\,L_\nu^0
           =10\,(\epB n_0)^{1/2}
           \,g_\nu\,\tm\,\tG^2\,\frac{(1+z)}{\pi D^2},
\ee
\be
   \tm=\frac{4\pi}{3}\,\tR^3\,n_0m_p.
\ee
We focus here on low frequencies $\nu<\nu_c,\nu_m$ and keep only the first 
two lines in the general expression for $g_\nu$ (eq.~\ref{eq:g}). 
Substituting equations~(\ref{eq:gm}) and (\ref{eq:gc}) for $\nu_m$ and 
$\nu_c$ and using $a\approx 1$ (\S~5.1), one gets
\begin{eqnarray}
\label{eq:g0}
  g_\nu \approx\left\{\begin{array}{ll}
  (\nu/\nu_m)^{1/3}=0.013\,\nu_{15}^{1/3}\,\tG_2^{-4/3}
                    (\epB n_0)^{-1/6}\psi^{-2/3}  & \nu<\nu_m<\nu_c, \\
  (\nu/\nu_c)^{1/3}=0.48\,\nu_{15}^{1/3} \tR_{16}^{2/3}
                    (\epB n_0)^{1/2} & \nu<\nu_c<\nu_m, 
                      \end{array}\right.
\end{eqnarray}
Note that $\nu=(1+z)\nuobs$ in all the formulae. Substitution of 
$\tR(\tobs)$ and $\tG(\tobs)$ gives $\Fnuobs^0$ as a function of $\tobs$.

In particular, at $\tobs<\tdec$ we have $\tG=\Gamma_0=const$ and get
\be
\label{eq:stand1}
  \frac{\Fnuobs^0}{\rm Jy}=\frac{0.3}{D_{28}^2}\,
      \left(\frac{\nuobs}{10^{15}{\rm Hz}}\right)^{1/3}
   \left\{\begin{array}{ll}
             10^{-9}\,\left(\frac{\Gamma_0}{100}\right)^{20/3}\,n_0^{4/3}
 \,\epB^{1/3}\,\psi^{-2/3}\,(1+z)^{-5/3}\,\tobs^3 & \nu<\nu_m<\nu_c,    \\
             10^{-8}\,\left(\frac{\Gamma_0}{100}\right)^{28/3}\,n_0^{2}
        \,\epB\,(1+z)^{-7/3}\,\tobs^{11/3} &  \nu<\nu_c<\nu_m.  
\end{array}\right.
\ee

At $\tobs>\tdec$, we have for an adiabatic blast wave 
$\tG^2\tm c^2=\Eej=const$, and rewrite $\Fnuobs$ as 
\be
   \Fnuobs= 0.3\; (1+z)\left(\frac{\Eej}{10^{53}\rm erg}\right)\,
            D_{28}^{-2}(n_0\epB)^{1/2}\,g_\nu \; {\rm Jy},
\ee
which gives
\be
\label{eq:stand2}
  \frac{\Fnuobs^0}{\rm Jy}=\frac{0.3}{D_{28}^2}\,
  \left(\frac{\nuobs}{10^{15}{\rm Hz}}\right)^{1/3}
   \left\{\begin{array}{ll}
             10^{-3}(\Eej/10^{53})^{5/6}\,n_0^{1/2}\,
   \epB^{1/3}\,\psi^{-2/3}\, (1+z)^{5/6}\,\tobs^{1/2} & \nu<\nu_m<\nu_c, \\
        (\Eej/10^{53})^{7/6}\,n_0^{5/6}\,
   \epB\,(1+z)^{7/6}\,\tobs^{1/6} &  \nu<\nu_c<\nu_m.  
\end{array}\right.
\ee
Here $\tobs$ is in seconds and $n_0$ is in cm$^{-3}$; $\psi$
is given by equation~(\ref{eq:psi}). Same results were derived previously 
(e.g. Sari et al. 1998; Panaitescu \& Kumar 2000).

\subsection{Pair-dominated part}

The characteristic frequency $\tnuacc(\tobs)$ is given by
\be
  \tnuacc=4\times 10^{14}\,\left(\frac{\Gamma_0}{100}\right)^2
  \left(\frac{\epsilon_{B\rm acc}n_0}{0.01}\right)^{1/2}
  \left(\frac{\psi}{0.1}\right)^2
              \left\{\begin{array}{ll}
    1                  & \tobs<\tdec \\
  (\tobs/\tdec)^{-9/8} & \tobs>\tdec 
                     \end{array}
              \right\} \, {\rm Hz}.
\ee
We have $f_\nu\sim const$ in a broad range of frequencies around $\tnuacc$ 
(see \S~5.2.2) and therefore
\be
\label{eq:homog1}
   L_\nu^\pm \approx const\; \tep_{B\rm acc}^{1/2}. 
\ee
The observed flux is given by 
\beq
\label{eq:F_homog1}
   \frac{\Fnuobs^\pm}{\rm Jy} =  \frac{\tG^2\,(1+z)}{2\pi D^2}\,L_\nu^\pm
                =  1.7\times 10^{2}\,(1+z)\,\frac{\Racc^3}{D^2}
              \,n_0^{3/2}\,\tep_{B\rm acc}^{1/2}\tG^2\;f_\nu,
\eeq
where we have used
\be
\label{eq:macc}
   \macc=\frac{4\pi}{3}\,\Racc^3\,n_0\,m_p,
\ee
and $\Racc$ is given by equation~(\ref{eq:Rxi}). Finally, we substitute
equation~(\ref{eq:G}) for $\tG(\tobs)$ and find
\beq
\nonumber
    \frac{\Fnuobs^\pm}{\rm Jy} 
                &=& 1.7\times 10^{-2}\,(1+z)\,
                     \left(\frac{\Gamma_0}{100}\right)^2\,
                     \left(\frac{\Racc}{10^{16}\rm ~cm}\right)^3\,
                     \left(\frac{D}{10^{28}\rm ~cm}\right)^{-2}\,
                     n_0^{3/2}\, \\
                & & \times f_\nu\,\tep_{B\rm acc}^{1/2} 
                     \left\{\begin{array}{ll}
                             1               & \tobs<\tdec, \\
                        (\tobs/\tdec)^{-3/4} & \tobs>\tdec.
                                \end{array}
                     \right.
\label{eq:F_homog}
\eeq
Equation~(\ref{eq:F_homog}) shows that before $\tdec$ the observed flux 
evolves as $\Fnuobs^\pm(\tobs)=const\,f_\nu\,\tep_{B\rm acc}^{1/2}$,
where $f_\nu$ changes very slowly while
$\tep_{B\rm acc}^{1/2}$ increases as $\tobs$ 
if the postshock magnetic flux is conserved (eq.~\ref{eq:tepB_homog}) and 
reaches $(\Rdec/\Racc)\epsilon_B^{1/2}$ at $\tobs=\tdec$. After $\tdec$, 
the observed flux decreases as $\tep_{B\rm acc}^{1/2}\tG^2f_\nu$ with 
$\tep_{B\rm acc}^{1/2}\approx const=(\Rdec/\Racc)\epsilon_B^{1/2}$; then 
\be
  \Fnuobs^\pm\propto\tG^2f_\nu\propto 
     \left(\frac{\tobs}{\tdec}\right)^{-3/4-(9/8)\alpha},
\ee
where $\alpha\sim 0.2$ is the slope of spectral function $f_\nu$,
and we get the approximate decay rate $\Fnuobs^\pm\propto \tobs^{-0.9}$. 
A faster decay of $\Fnuobs^\pm$ is possible only if the magnetic field is 
destroyed in the blast wave and $\epsilon_{B\rm acc}$ decreases with $\tR$.

\subsection{Radiative cooling of $e^\pm$}

The above analytical calculation of $F_\nu^\pm$ assumed that the 
pairs are slowly cooling. We now check this assumption.

\subsubsection{The cooling cutoff of the pair distribution}

The radiative cooling defines the cutoff of the $e^\pm$ distribution 
$\gamma_c(m,\tR)$ (\S~3.3) and the corresponding cutoff frequency of the 
synchrotron spectrum $\nu_c(m,\tR)$,
\be
   \nu_c(m,\tR)=0.2\tG\frac{e\tB}{m_ec}\,\gamma_c^2(m,\tR),
\ee 
where $\tG$ is the current Lorentz factor of the blast wave and 
$\tB=B(m,\tR)$. The radiative cooling does not affect $L_\nu^\pm(\tR)$ 
if $\gamma_c(m,\tR)>\gamma_m(m,\tR)$ at $m\sim\macc$.

Both synchrotron and inverse Compton losses can affect the value of $\gamma_c$
at $\macc$. They are proportional to the magnetic energy density 
$U_B(\macc,\tR)$ and soft radiation density $U_s(\macc,\tR)$, respectively. 

As the blast wave expands from $\Racc$ to some $\tR<\Rdec$, its energy 
density $\tU$ remains approximately constant while $\tep_{B\rm acc}$ grows 
as $(\tR/\Racc)^2$ (assuming that the magnetic flux is conserved, see 
eq.~\ref{eq:tepB_homog}), so $U_B(\macc,\tR)\approx (\tR/\Racc)^2\epB\tU$. 
$U_s(\macc,\tR)$ is dominated by synchrotron radiation transported from the 
forward shock, which is in the fast-cooling regime. The density of 
synchrotron radiation at the current position of the shock is 
$U_s(\tm,\tR)\approx (\epe\epB)^{1/2}\tU$ (the rest of electron energy is 
taken away by inverse Compton scattering of synchrotron radiation, e.g. 
Sari \& Esin 2001). A fraction $\sim 0.2$ of this radiation is transported 
to the downstream region of the blast wave.\footnote{The transport 
coefficient is evaluated assuming isotropic emission from the forward shock 
in the blast-wave frame. Its dependence on the downstream Lagrangian 
coordinate $m$ is weak as long as $m\ll\tm$. For example, the contact 
discontinuity ($m=0$) receives $\sim 0.15$ of the forward-shock emission.} 
Then the density of soft radiation at $m\sim\macc$ is 
\be
\label{eq:U_s}
 U_s(\macc,\tR)\approx 0.2\,\left(\epe\epB\right)^{1/2}\tU.
\ee
This implies that $U_s(\macc,\tR)$ is approximately constant $\tR<\Rdec$.

So, we find that the relative contribution of inverse Compton scattering to 
the cooling rate at $m\sim\macc$ is given by
\be
  C(\macc,\tR)=\frac{U_s(\macc,\tR)}{U_B(\macc,\tR)}
  \approx 0.2\,\frac{(\epe\epB)^{1/2}}{\tep_{B\rm acc}}
 =0.2\,\left(\frac{\Racc}{\tR}\right)^2\left(\frac{\epe}{\epB}\right)^{1/2},
\ee
where the last equality assumes conservation of magnetic flux. Below
we consider blast waves with $C(\macc,\Rdec)\simlt 1$; then the cutoff 
at $m\sim\macc$ is shaped by synchrotron losses rather than inverse 
Compton scattering of the forward-shock radiation.

The cooling peaks at the deceleration radius and shapes the cutoff of the
$e^\pm$ spectrum,
\beq
\nonumber
  \gamma_c(m,\Rdec)&=&\frac{3m_e}{16\,m_p\,\sT\,\tepB n_0\,\tG\,(\Rdec-R)} \\
  &=&\frac{9.62\times 10^{-4}\epB^{-1}\Gamma_0}
        {\left[\gamma(1+\beta)\right]^{1/2}}
       \,\left(\frac{R}{10^{16}\,\rm cm}\right)^2
         \left(\frac{\Eej}{10^{53}\,\rm erg}\right)^{-1}
         \left(1-\frac{R}{\Rdec}\right)^{-1},
\eeq
\beq
\nonumber
   \nu_c(m,\Rdec)&=& 3.25\times 10^{15}
   \left[\frac{\epsilon_B(m,\Rdec)\;n_0}{0.01}\right]^{-3/2}
   \left(\frac{\Rdec}{10^{17}\rm ~cm}\right)^{-2}
   \left(1-\frac{R}{\Rdec}\right)^{-2} \; {\rm Hz} \\
  &=& \frac{3.25\times 10^{15}}{[\gamma(1+\beta)]^{3/4}}
   \left(\frac{\epB\,n_0}{0.01}\right)^{-3/2}\left(\frac{m}{\mdec}\right)
   \left(\frac{\Rdec}{10^{17}\rm ~cm}\right)^{-2}
   \left(1-\frac{R}{\Rdec}\right)^{-2} {\rm Hz}.
\eeq
where $R\approx\Racc$ is the radius where the pair-loaded mass 
$m\sim\macc$ was shocked and $\gamma(1+\beta)$ is the preacceleration 
factor at this radius. The last equalities in these expressions
assume magnetic flux conservation. In particular, the factors $m/\mdec$
and $[\gamma(1+\beta)]^{-3/4}$ appeared in the expression for 
$\nu_c(m,\Rdec)$ because of the flux conservation; they would be absent if
$\epB(m,\tR)=const$. 

When the blast wave reaches the deceleration radius, its pressure starts 
to decrease, and the afterward evolution of the $e^\pm$ distribution 
function at $m\sim\macc$ is fully determined by adiabatic cooling --- 
radiative cooling is slower and has a negligible effect. The adiabatic 
cooling of the blast wave expanded from $\Rdec$ to a current $\tR$ is 
described by the factor $A=(\tG/\Gamma_0)^{1/2}$ (\S~3.3), and the resulting 
cutoff Lorentz factor is 
\be
 \gamma_c(m,\tR)=\left(\frac{\tG}{\Gamma_0}\right)^{1/2} 
                \gamma_c(m,\Rdec), \qquad \tR>\Rdec.
\ee
The corresponding cutoff frequency evolves as $\tnu_c\propto \tG\tB\tg_c^2$, 
which gives 
\be
   \frac{\nu_c(m,\tR)}{\nu_c(m,\Rdec)}=\left(\frac{\tG}{\Gamma_0}\right)^3
   \left[\frac{\epsilon_B(m,\tR)}{\epsilon_B(m,\Rdec)}\right]^{1/2},
   \qquad \tR>\Rdec.
\ee
It can be expressed as a function of $\tobs$ using equation~(\ref{eq:G}) 
for $\tG$ and a prescription for the magnetic field evolution in the blast
wave. If the magnetic flux is conserved, $\tepB^{1/2}$ evolves 
as $\tobs^{1/16}$ (eq.~\ref{eq:tepB_homog}), and we get,
\be
   \frac{\nu_c(m,\tobs)}{\nu_c(m,\tdec)}
       =\left(\frac{\tobs}{\tdec}\right)^{-9/8+1/16}
    \approx \left(\frac{\tobs}{\tdec}\right)^{-17/16},
    \qquad \tobs>\tdec.
\ee
If $\epB$ remained constant at all $m$ and $\tR$, one gets the cutoff
frequency $\nu_c(m,\tobs)\propto \tobs^{-9/8}$, almost the same as with 
flux conservation.

\subsubsection{The slow-cooling condition}

A given shell $m$ in a blast wave of radius $\tR$ is in the slow cooling 
regime if $\gamma_c(m,\tR)>\gamma_m(m,\tR)$. Using equation~(\ref{eq:gm_}),
\be
 \gamma_m(m,\tR)=\frac{\Gamma_0}{Z\gamma(1+\beta)}\,\psi\,\frac{m_p}{m_e},
\ee
we find with the flux conservation assumption,
\be
\label{eq:ratio_nu}
   \left[\frac{\nu_m(m,\Rdec)}{\nu_c(m,\Rdec)}\right]^{1/2}=
   \frac{\gamma_m(m,\Rdec)}{\gamma_c(m,\Rdec)}=
    \frac{1.91\times 10^{6}\,\epB\,\psi}{Z\,[\gamma(1+\beta)]^{1/2}}
    \left(\frac{R}{10^{16}\,\rm cm}\right)^{-2}
    \left(\frac{\Eej}{10^{53}\,\rm erg}\right)
    \left(1-\frac{R}{\Rdec}\right).
\ee
Note that $n_0$ and $\Gamma_0$ cancel out from this expression.
This ratio remains the same at $\tR>\Rdec$ as both $\gamma_m(m,\Rdec)$ 
and $\gamma_c(m,\Rdec)$ evolve adiabatically and are reduced by a common 
factor $A$. The effect of the $\gamma$-ray precursor enters through the 
factor $Z^{-1}[\gamma(1+\beta)]^{-1/2}$, which reduces the ratio.

We are interested in $m\approx\macc$ because $L_\nu^\pm$ peaks near $\macc$.
Equation~(\ref{eq:ratio_nu}) shows that the slow-cooling assumption is 
valid if
\be
   \frac{1.91\times 10^{6}\,\epB\,\psi}{\Zacc}
    \left(\frac{\Racc}{10^{16}\,\rm cm}\right)^{-2}
    \left(\frac{\Eej}{10^{53}\,\rm erg}\right)<1,
\ee
which can be rewritten as 
\be
  \epB\epe<4\times 10^{-5}\left(\frac{p-1}{p-2}\right)
     \left(\frac{\Racc}{10^{16}\rm cm}\right)^{2}
     \left(\frac{\Eej}{10^{53}\rm erg}\right)^{-1}
 \approx 2\times 10^{-5}\left(\frac{p-1}{p-2}\right)
                \left(\frac{E_\gamma}{\Eej}\right).
\ee


\section{Comparison with the reverse-shock model}

We focused in this paper on the forward shock of the GRB explosion.
Early optical radiation is also expected from the reverse-shock in the 
ejecta if their energy is dominated by baryons rather than magnetic 
field (M\'esz\'aros \& Rees 1993; Sari \& Piran 1999). 
Explosions with the reverse and $e^\pm$-loaded forward shocks should 
have two soft emission components, and it is instructive to compare them.

Emission from the reverse shock peaks when it crosses the ejecta, which 
occurs at the deceleration radius. Then the explosion Lorentz factor 
$\tG$ decreases and the reverse-shock emission decays. The decay rate can 
be calculated as follows. The Lorentz factor of electrons emitting at 
a fixed frequency $\nuobs=(1+z)\nu$ is
\be
  \gamma_e=\gamma_\nu=\left[\frac{5\nu m_ec}{\tG\tBej}\right]^{1/2}
  \propto(\tBej\tG)^{-1/2},
\ee
where $\tBej$ is the magnetic field in the ejecta at the current radius 
$\tR$. The number of electrons emitting at frequency $\nu$ is 
\be
  N_\nu\approx N_e(\gamma_\nu/\gamma_m)^{-p+1},
\ee
where $N_e$ is the total number of shock-accelerated electrons and 
$\gamma_\nu>\gamma_m$ is assumed. With decreasing $\tG$, $N_\nu$ is 
reduced for two reasons: (1) $\gamma_\nu$ increases and (2) $\gamma_m$ 
decreases as the whole nonthermal spectrum is shifted to lower energies 
by adiabatic cooling. The adiabatic cooling factor is 
$A\propto\tP^{1/4}$ where $\tP=(4/3)\tG^2\tn_0\mu_em_pc^2$ is the 
blast-wave pressure, and hence
\be
  N_\nu\propto \gamma_\nu^{-p+1}A^{p-1}
       \propto \left(\tBej\tG\right)^{(p-1)/2}\tn_0^{(p-1)/4}\tG^{(p-1)/2}. 
\ee
The observed synchrotron flux is given by
\be
  \Fnuobs^{\rm RS}=\frac{5(1+z)}{36\pi^2 D^2}\,\frac{m_ec^2\,\sT}{e}\,
     \tBej\tG\,N_\nu\propto\tn_0^{(p-1)/4}\,\tBej^{(p+1)/2}\,\tG^p.
\ee
If no destruction of magnetic flux takes place, $\tBej$ evolves as 
\be
  \tBej\propto\tn_{\rm ej}\tR\propto P^{1/\hat{\gamma}}\tR, 
\ee
where $\hat{\gamma}$ is the adiabatic index of the ejecta material. Then,
\be
  \Fnuobs^{\rm RS}\propto \tn_0^{(p+1)/2\hat{\gamma}+(p-1)/4}\,
  \tG^{(p+1)/\hat{\gamma}+p}\,\tR^{(p+1)/2}.
\ee
In the case of a uniform ambient medium, $\tn_0=const$, we have
after the deceleration radius $\tG\propto\tobs^{-3/8}$ and 
$\tR\propto\tobs^{1/4}$, which yields
\be
  \Fnuobs^{\rm RS}\propto\tobs^{-3(p+1)/8\hat{\gamma}-p/4+1/8}
   =\left\{\begin{array}{ll}
      \tobs^{-1.3}, & \hat{\gamma}=5/3,\; p=2.5 \\
      \tobs^{-1.5}, & \hat{\gamma}=4/3,\; p=2.5 \\
           \end{array}
    \right.
\ee
A destruction of magnetic field in the ejecta could only steepen the decay.

The mechanism of reverse-shock emission is similar to that of the $e^\pm$
afterglow: a shell of material (ejecta in the case of reverse shock and 
$\macc$ in the case of $e^\pm$ loaded forward shock) is heated at $R<\Rdec$ 
with a low energy per particle, and after $\Rdec$ the shell cools down 
passively (adiabatically) producing a decaying flux of soft emission.

The main difference between the two cases is the spectrum of emitting 
particles. The reverse-shock is thought to produce a power-law electron
distribution with $p=2-3$. By contrast, the effective spectrum of $e^\pm$ 
has $p\approx 1$. In the narrow shell $\Delta m\sim \macc$, there are 
approximately equal numbers of $e^\pm$ at all energies up to the cooling 
cutoff, which gives the effective $p$ about unity. The flat distribution 
of $e^\pm$ leads to the slow decay of their synchrotron emission that we 
found in \S~6. The resulting light curve can, however, be made steeper if 
the ambient density decreases with radius or magnetic field is gradually 
destroyed downstream of the shock. Therefore, the main intrinsic difference 
between the reverse-shock and $e^\pm$ emission components is not the 
produced light curve but the slope of the $e^\pm$ distribution $p$. 

This difference can be observed directly by measuring the instantaneous 
synchrotron spectrum. The $e^\pm$ radiation is expected to have a small 
spectral index $|\alpha|<0.2$ while the reverse-shock spectrum is much 
steeper: $\alpha=(p-1)/2=0.5-1$ for $p=2-3$. A measurement of the 
instantaneous spectrum in UV, optical, or IR at $\tobs\sim 100$~s would 
provide a test of the current theoretical picture of the GRB explosion. 
Such a test can be done by {\it Swift}.

Finally, we note one more difference between the reverse-shock and $e^\pm$ 
emission components. Both are cut off when $\nu_c<\nu$, which happens at 
different times because the cooling frequencies $\nu_c$ are different in 
the two cases. The ejecta magnetic field is likely stronger, $\Bej>B$; 
then its cooling frequency is lower by the factor $(\Bej/B)^{-3}$ and 
hence the cutoff should occur sooner.


\section{Discussion}

\subsection{$e^\pm$ component of GRB afterglow}

An explosion blast wave is composed of swept-up layers (shells) of external 
medium that have been shocked in the forward shock front. In GRB explosions, 
these layers contain $e^\pm$ pairs injected into the external medium by the 
$\gamma$-ray front. The layers shocked at small radii $R<\Rload$ are 
dominated by the numerous nonthermal $e^\pm$. At $R>\Rload$, these layers 
remain in the blast wave and form a thin $e^\pm$ shell adjacent to the 
contact discontinuity if there is no turbulent mixing in the blast. This 
shell has a low energy per particle and emits much softer synchrotron 
radiation compared to the outer swept-up material. The optical emission of 
the expanding blast wave can be strongly dominated by the $e^\pm$ shell 
even at large radii where its mass $\mload$ is small compared to the total 
swept-up mass. If the shock magnetic parameter $\epB$ is below a critical 
value $10^{-4}-10^{-3}$, the optical-emitting $e^\pm$ are in the 
slow-cooling regime and radiate their energy slowly, on a timescale longer 
than the deceleration time of the explosion. Their effect on the observed 
afterglow is described by adding a new emission component $\Fnuobs^\pm$.

The $e^\pm$ afterglow component is less sensitive to the model assumptions 
than the customary pair-free afterglow. This is because the $e^\pm$ shell 
contains approximately equal numbers of particles with vastly different 
energies, which is a consequence of the steep evolution of the $\gamma$-ray
front and $e^\pm$ loading with radius. This fact allowed us to derive a 
simple formula for the observed spectral flux,
\be
\label{eq:fl}
   \Fnuobs^\pm=6\times 10^{-7}\,\tG^2\frac{(1+z)}{D_{28}^2}\,
   \tep_{B\rm acc}^{1/2} n_0^{3/2}
   \left(\frac{E_\gamma}{10^{53}\,\rm erg}\right)^{3/2} f_\nu \;{\rm Jy}.
\ee
It depends on the blast-wave Lorentz factor $\tG$ that approximately
corresponds to a given observer time $\tobs$. In a broad range of $\nu$,
the flux is dominated by a specific mass shell $m\approx\macc$ whose shock 
radius was $\Racc=7\times 10^{15}E_{\gamma,53}^{1/2}$~cm. $\Fnuobs^\pm$ 
depends on the magnetic field in this mass shell, $\tB_{\rm acc}$, which 
we parameterize using the usual equipartition parameter $\tep_{B\rm acc}$ 
($\tB_{\rm acc}^2/8\pi$ divided by the energy density of the blast wave).
Accurate calculations in \S\S~4 and 5 give the correction factor 
$f_\nu\sim 1$ in equation~(\ref{eq:fl}). All the details and uncertainties 
of the blast wave physics are absorbed by this factor (which is never much 
different from unity) and $\tep_{B\rm acc}$.

To the first approximation, $\Fnuobs$ decays after the deceleration time 
as $\tG^2$ which is proportional to $\tobs^{-3/4}$ for an adiabatic blast 
wave in a uniform medium. The decay can, however, be faster if 
$B_{\rm acc}$ is gradually destroyed as the blast wave expands.

\subsection{Uncertainties in the blast wave physics}

\subsubsection{Magnetic field downstream of the shock}

The magnetic field behind the shock front and its downstream evolution
are a major uncertainty of the current blast-wave models.
In this paper we assumed, as is customary, that the magnetic parameter 
$\epB$ immediately behind the shock front remains constant as the shock 
expands, and followed the downstream evolution of $\epsilon_B(m,\tR)$ 
assuming conservation of the postshock magnetic flux. This assumption 
unambiguously determines the magnetic field in the $e^\pm$ shell, and we 
found that the shell is in the slow-cooling regime if the shock parameter 
$\epB$ is below a critical value $10^{-4}-10^{-3}$.
If $\epB$ is above this value, the $e^\pm$ shell is in the 
fast-cooling regime and most of its energy is radiated at $\tobs<\tdec$.
In this case, the $e^\pm$ radiation can be visible at $\tobs>\tdec$
only in the infrared band; the time of the emission cutoff depends on 
frequency as $\nuobs^{-1}$.

The results are qualitatively similar if $\epB$ does not evolve in the
postshock region according to the flux conservation but instead 
remains constant (which requires a gradual destruction of the magnetic 
flux with increasing distance behind the shock).
However, the results would change if the magnetic flux is destroyed more 
quickly. The luminosity of the $e^\pm$ shell is proportional to its 
magnetic field and the field destruction would give a faster decay of 
$\Fnuobs^\pm$. 

A real postshock field can be inhomogeneous on small scales (Medvedev \& 
Loeb 1999) and one may need to include a distribution $f(\epB)$ in future 
afterglow models. This can change details. For instance, the cooling cutoff
would be less pronounced because the low-$\epB$ fraction would continue 
to radiate in the slow-cooling regime and give a tail of emission even 
when most of the postshock plasma is cooled and does not emit at the 
observed frequency.

\subsubsection{Turbulent mixing}

The $e^\pm$-dominated material may not form a distinct thin shell near
contact discontinuity if there are large-scale turbulent motions that 
mix up the postshock layers. Rayleigh-Taylor instability can drive such 
mixing like it does in supernova remnants. The mixing is unlikely to 
change the results of this paper. As long as we approximate the blast wave 
as a constant-pressure shell, the pressure in an $e^\pm$-dominated gas 
element does not depend on its position within the blast wave.
Therefore, its adiabatic and radiative cooling is the same as in the 
absence of turbulent mixing, and conservation of magnetic flux gives 
the same magnetic field in the element.

\subsubsection{Mechanism of electron acceleration}

A significant uncertainty in the afterglow physics is the mechanism of
electron acceleration. A preshock magnetic field is compressed in the 
relativistic shock and becomes transverse, and the fields generated by 
Weibel instability are also transverse --- parallel to the shock plane 
(Medvedev \& Loeb 1999). Standard diffusive acceleration is unlikely to be 
efficient under such conditions. It requires the electron to cross the shock 
front many times, which can hardly happen: the upstream diffusion across
the transverse field is slow because the electron gyroradius is smaller 
than the front thickness (proton gyroradius), and the electron will be 
advected downstream with the flow velocity $c/3$ with respect to the front 
before it gets a chance to diffuse back to the upstream region.

An alternative mechanism is stochastic acceleration by turbulence downstream
of the shock. It may not be well described as impulsive acceleration and 
may keep electrons energetic even in the presence of significant radiative or 
adiabatic losses. This could change the theoretical afterglow light curve.

However, the main signature of $e^\pm$ emission --- white spectrum ---
will likely persist because the blast wave will still have a steep 
variation of $e^\pm$ density with the Lagrangian coordinate $m$, which 
invariably leads to the broad $e^\pm$ distribution with $p\approx 1$.
This special feature ultimately comes from the $\gamma$-ray front evolution
with radius and is not related to the mechanism of $e^\pm$ acceleration.

\subsubsection{Electron distribution function}

Customary afterglow calculations assume an idealized distribution function 
of the postshock electrons: all electrons reside in a power-law component 
that starts at $\gamma_m$ and ends at $\gamma_c$. In reality, it is 
possible that only a small fraction of electrons $\zeta_e$ are accelerated 
in a shock wave and the rest of them form a quasi-Maxwellian distribution; 
this is observed to be the 
case for collisionless shocks in the solar system. The expected energy 
of the accelerated electron population $\epe$ is then typically 
$\sim 1$\% of the total plasma energy (which is dominated by the hot ions). 
Similar electron acceleration was inferred for supernova shocks. 
By contrast, $\epe$ inferred from the existing fits of GRB afterglows is 
$\epe\sim 0.1$ (e.g. Panaitescu \& Kumar 2002).

We point out that the high $\epe$ may have been inferred because the fits
assume $\zeta_e=1$. Equally good fits may be obtained with a more reasonable 
$\zeta_e\sim 0.1$ and $\epe\sim 0.01$. This can be understood by looking at 
how the parameters enter the emission model. Besides the spectral slope $p$,
the emitting electrons are described by two parameters: the number of 
accelerated particles $N_e=\zeta_e N_t$ (where $N_t$ is the total number of 
swept-up particles) and the minimum nonthermal Lorentz factor $\gamma_m$. 
Note that $\gamma_m m_ec^2$ is comparable with the mean energy per electron,
which is proportional to $\epe/\zeta_e$. Relaxing the assumption $\zeta_e=1$, 
one can get the same $\gamma_m$ by decreasing $\epe\propto\zeta_e$, and 
then the same observed emission may be explained with 10 times lower, and 
physically more plausible, values of $\zeta_e$ and $\epe$. 
The number of emitting particles $N_e$ will not be changed if $N_t$ is 
increased by the factor $\zeta_e^{-1}$. Thus, the reduction of $\zeta_e$ 
implies a higher ambient density and the circumburst density may have been 
systematically underestimated in the afterglow fits by one order of magnitude.

Different assumptions concerning the shape of electron distribution 
function can lead to different afterglow radiation. However, the $e^\pm$ 
afterglow component is almost insensitive to such assumptions. For 
illustrative purpose, consider an extreme case. Suppose there is no 
power-law acceleration at the shock front, and the shock produces a narrow 
$e^\pm$ distribution peaking at 
$\gamma_e(R)\approx\epe(m_p/m_e)\Gamma(\gamma Z)^{-1}$. The factor 
$\gamma Z$ is determined by the $\gamma$-ray transfer through the ambient 
medium and has a robust steep dependence on $R$ near $\Racc$ (\S~2). 
Therefore, $\gamma_e$ will depend steeply on $R$, and the resulting $e^\pm$ 
distribution in the shell $\Delta m\sim\macc$ swept-up near $\Racc$ 
will be broad and flat. The same formula (eq.~\ref{eq:fl}) will describe 
the $e^\pm$ radiation with a slightly different numerical factor 
$f_\nu\sim 1$.

\subsection{Prospects of detection of $e^\pm$ emission}

The $e^\pm$ radiation is predicted to have a special 
feature that can be tested with upcoming observations:
its spectral index $\alpha$ is close to zero ($|\alpha|\simlt 0.2$).
This is a significant difference from the reverse-shock model which 
predicts a steeper spectrum, $\alpha=(p-1)/2$, where $p=2-3$ is a
putative slope of the electron distribution. Although the early optical 
emission has already been caught in a few bursts, no spectral data are 
presently available. {\it Swift} can provide the valuable spectral 
information.

The $e^\pm$ dominance of the early UV/optical/IR afterglow can 
result in a characteristic two-peak shape of the light curve (see 
an example in Fig.~3). The $e^\pm$ emission component begins to decay at 
the deceleration time (eq.~\ref{eq:tdec}), which can be before {\it Swift} 
detects the afterglow. However, its decay is relatively slow 
($\Fnuobs^\pm\propto\Gamma^2\propto\tobs^{-3/4}$ in the first approximation)
and its tail is observable on timescales of minutes until the pair-free 
component takes over. The $e^\pm$ radiation should be visible for 
a longer time in the infrared band. 

To date, early optical emission ($\tobs<1000$~s) has been detected in 
four bursts: GRB~990123, GRB~021004, GRB~021211, and GRB~030418.
In only one of them, GRB~990123, the peak of the optical flash was 
caught (Akerlof et al.~1999). This peak overlapped with the prompt MeV 
burst, and hence the model developed in the present paper does not apply 
to GRB~990123: the $e^\pm$ must be Compton cooled by the MeV photons 
(keV in the fluid frame) and most of the $e^\pm$ energy is likely emitted 
in the GeV-TeV band (Beloborodov 2005). The strong optical $e^\pm$ radiation 
is expected in bursts where the MeV radiation front completely overtakes 
the blast wave by the time it reaches $\Racc$ as discussed in \S~1.

\subsection{Neutron front}

We studied in this paper the pair-loading effects on the forward shock.
If the ejecta contains baryons, i.e. is not a pure electro-magnetic outflow 
(Poynting flux), a significant fraction of the baryons must be free 
neutrons (Derishev, Kocharovsky, \& Kocharovsky 1999; Beloborodov 2003a). 
The neutron ejecta get completely decoupled and coast freely by the beginning 
of afterglow emission with a Lorentz factor $\Gamma_n\sim\Gej$. They gradually 
$\beta$-decay, however, some neutrons survive till radii $R\sim 10^{17}$~cm 
when the blast wave may have already decelerated, overtake the blast wave, 
and decay ahead of it depositing significant momentum and energy into the 
ambient medium (Beloborodov 2003b). Thus, GRB explosions are likely to 
develop leading neutron fronts that change the mechanism of the blast wave. 

The neutron front may emerge either after the $e^\pm$-loading, at 
$R>\Rload$, or at smaller radii, depending on the presence of fast 
neutrons with $\Gamma_n>\Gamma$. The impact of a fast neutron 
front on the early afterglow was recently studied by Fan, Zhang, \& Wei 
(2004), and an alternative scenario with slow neutrons was proposed by 
Peng, K\"onigl, \& Granot (2004). Accurate afterglow calculation that 
includes both the neutron decay and $e^\pm$ loading is a challenging 
problem which can be solved in future. We expect that the main signature 
of $e^\pm$-loading --- soft emission with a broad flat spectrum --- be 
present in neutron-fed afterglows as well.

\acknowledgments
I am grateful to Frederic Daigne, Robert Mochkovitch, and Chris Thompson 
for discussions, and the IAP for hospitality. I also thank Jules Halpern
and Davide Lazzati, the referee, for comments on the manuscript. This work 
was supported by NASA grant NAG5-13382 and Alfred P. Sloan Fellowship.


\section*{Appendix: Calculation of $L_\nu^\pm$ integral}

We here calculate the integral $L_\nu^\pm$ neglecting the cooling cutoff 
of the $e^\pm$ distribution function (i.e. assuming $\tnu_c>\nu,\tnu_m$). 
Then,
\be
\label{eq:dLdm}
   \frac{\delta L_\nu}{\delta m}=\frac{\dLmax_\nu}{\delta m}\times
    \left\{\begin{array}{ll}
      (\nu/\tnu_m)^{1/3} & \tnu_m\geq\nu, \\
      (\nu/\tnu_m)^{(1-p)/2} & \tnu_m\leq\nu, \\
           \end{array}
    \right.
\ee
\be
\label{eq:dLmaxdm}
   \frac{\dLmax_\nu}{\delta m}(m,\tR)
         =w(\tR)\,\tepB^{1/2}\frac{Z}{\mu_e},
   \qquad w(\tR)=30\,\left[\frac{\tn_0\mu_e}{\tg(1+\tb)}\right]^{1/2} 
   \; \frac{\rm erg}{\rm g}.
\ee
We have written $\dLmax_\nu/\delta m$ in this form to separate quantities
that do not depend on the Lagrangian coordinate (functions of the
current radius $\tR$ only) and are constant in the $L_\nu^\pm$ integral.
In a similar way, we rewrite the expression for $\nu_m(m,\tR)$ 
(eq.~\ref{eq:tnu_m}),
\be
\label{eq:tnu_m_}
 \nu_m(m,\tR)=q(\tR)\,\frac{\tepB^{1/2}\,\Gamma}
  {n_0^{1/2}[\gamma(1+\beta)]^{3/2}(Z/\mu_e)^2}, \qquad 
  q(\tR)=4.6\times 10^{12}\frac{\tG^3\tn_0\mu_e^{1/2}\psi^2}{\tg(1+\tb)}
  \; {\rm Hz}.
\ee

Since $\tnu_m(m)=\nu_m(m,\tR)$ is a monotonic function of the Lagrangian
coordinate $0<m<m_1$, we can change the integration variable to $\nu_m$,
\be
  L_\nu^\pm=\int_0^{m_1}\frac{\delta L_\nu}{\delta m}\,\dd m
  =\int_0^{\tnu_1}\frac{\delta L_\nu}{\delta m}\,
  \left(\frac{\partial \nu_m}{\partial m}\right)_{\tR}^{-1}\,\dd\nu_m,
\ee
where $\tnu_1$ is the peak frequency of synchrotron emission at $m_1$,
\be
  \label{eq:tnu_1}
   \tnu_1= \left\{\begin{array}{ll} 
  \nu_m(\tm,\tR)=4.6\times 10^{12}\,\frac{\tG^4\tn_0^{3/2}}
  {[\tg(1+\tb)]^{5/2}\tilde{Z}^2}\,\epB^{1/2}\psi^2\mu_e^{5/2} {\rm ~Hz} &
    \tR\leq\Rload, \\
  \nu_m(\mload,\tR) & \tR\geq\Rload. \\
\end{array}
    \right.
\ee
The partial derivative of $\nu_m$ can be written as
\be
\label{eq:der}
  \left(\frac{\partial \nu_m}{\partial m}\right)_{\tR}
  =\frac{\nu_m}{m}\, \frac{\partial \ln\nu_m}{\partial \ln m}
  =\frac{\nu_m}{m}\, \left[  \frac{\partial}{\partial\ln m}
    \ln\left(\frac{\tepB^{1/2}\Gamma}{n_0^{1/2}(1+\beta)^{3/2}}\right)
   -\frac{\partial}{\partial\ln m}\left(\frac{Z^2}{\mu_e^2}\,
    \gamma^{3/2}\right) \right].
\ee
$Z^2\gamma^{3/2}$ varies with $m$ much faster than
$\tepB^{1/2}\Gamma n_0^{-1/2}(1+\beta)^{-3/2}$ (in the latter, only 
$\Gamma$ could vary significantly, but even that does not happen for
explosions with $\Rdec>\Racc$). Therefore, the second term in 
equation~(\ref{eq:der}) is dominant and the first term can be neglected. 
Then,
\be
  \frac{\partial \ln\nu_m}{\partial\ln m}
  \approx -\frac{\partial}{\partial\ln m}
   \left(\frac{Z^2}{\mu_e^2}\, \gamma^{3/2}\right)
 =-\frac{\partial \ln R}{\partial\ln m}\,\frac{\partial \ln\xi}{\partial\ln R}
 \,\frac{\dd}{\dd\ln\xi}\left(\frac{Z^2}{\mu_e^2}\,\gamma^{3/2}\right)
 =-\frac{2}{k}\,\frac{\dd}{\dd\ln\xi}
  \left(\frac{Z^2}{\mu_e^2}\,\gamma^{3/2}\right).
\ee
$Z/\mu_e$ and $\gamma$ are given as functions of $\xi$ in 
equations~(\ref{eq:Z}) and (\ref{eq:gam}), and we find 
\be
\label{eq:nu_der}
  \left(\frac{\partial \nu_m}{\partial m}\right)_{\tR}
    =\frac{\nu_m}{m}\,\frac{s}{k}\times\left\{\begin{array}{ll}
    10\,\xi(m)/\xiacc & m\geq\macc, \\
    17/2 & m\leq\macc. \\
        \end{array}
    \right.
\ee

Using equations~(\ref{eq:dLdm}) and (\ref{eq:dLmaxdm}) we get
\be
\label{eq:integral}
  L_\nu^\pm(\nu,\tR)=w(\tR)\int_0^{\tnu_1} \tepB^{1/2}\frac{Z}{\mu_e}
  \left\{\begin{array}{ll}
      (\nu_m/\nu)^{-1/3} & \nu_m\geq\nu \\
      (\nu_m/\nu)^{(p-1)/2} & \nu_m\leq\nu \\
           \end{array}
    \right\}
  \left(\frac{\partial \nu_m}{\partial m}\right)_{\tR}^{-1}\,\dd\nu_m.
\ee
$Z$ needs to be expressed as a function of $\nu_m$.
From equations~(\ref{eq:Z}) and (\ref{eq:gam}),
\beq
   \gamma= \left\{ \begin{array}{ll}
  1 & Z/\mu_e\leq e^5/2\approx 74, \\
  (2Z/e^5\mu_e)^{3/2} & Z/\mu_e\geq e^5/2, \\
                   \end{array}
           \right.
\eeq
which we substitute into equation~(\ref{eq:tnu_m_}) and then
express $Z$ from that equation,
\be
\label{eq:Z_nu}
  \frac{Z}{\mu_e}=\left\{ \begin{array}{ll}
  \left[\nu_m(1+\beta)^{3/2}n_0^{1/2}\,q^{-1}\tepB^{-1/2}\Gamma^{-1}
  \right]^{-1/2}               & m(\nu_m)\geq\macc,\\
  \left[\nu_m(1+\beta)^{3/2}n_0^{1/2}\,q^{-1}\tepB^{-1/2}\Gamma^{-1}
  \right]^{-4/17} 74^{9/17}    & m(\nu_m)\leq\macc.\\
                   \end{array}
           \right.
\ee
The quantities $n_0$, $\tepB$, and $\Gamma$ vary with $m$ and 
therefore vary with $\nu_m$ when $\nu_m$ is chosen as the independent 
variable. However, their variation with $\nu_m$ is slow
(because $\nu_m$ is a steep function of $m$). Therefore, with
sufficient accuracy, we have $Z\propto \nu_m^{-1/2}$ at $m(\nu_m)>\macc$
and $Z\propto \nu_m^{-4/17}$ at $m(\nu_m)<\macc$.

Now the integral~(\ref{eq:integral}) can be calculated.
We consider first $\tR\leq\Racc$ and then $\tR\geq\Racc$.

\subsection*{$\tR\leq\Racc$}

At radii $\tR<\Racc$ the Lagrangian coordinate $m\leq\tm<\macc$; then
$\partial\ln\nu_m/\partial\ln m=(17/k)$ and 
$Z\propto \nu_m^{-4/17}$.

For $\nu>\tnu_1$ we have $\nu_m<\nu$ in the whole 
blast, and the integral~(\ref{eq:integral}) reads
\beq
\nonumber
 L_\nu^\pm(\nu,\tR) 
    = w(\tR)\,\frac{k}{2}\,\epsilon_B^{1/2}\,\frac{Z_1}{\mu_e}\,m_1
      \,\left(\frac{\nu}{\tnu_1}\right)^{(1-p)/2}\,\frac{4}{(17p-25)}
  \qquad \nu\geq\tnu_1.
\label{eq:integ1}
\eeq
where $Z_1\equiv Z(m_1)$ and we have used $Z/Z_1=(\nu_m/\tnu_1)^{-4/17}$. 
We also used the fact that $\tepB^{1/2}$ and $m$ vary slowly with $\nu_m$ 
and took them as constants evaluated at $\nu_m=\tnu_1$ where the integral 
peaks.

For $\nu<\tnu_1$ the integral can be written as a sum of two integrals
over $\nu_m<\nu$ and $\nu_m>\nu$. Both integrals peak at $\nu_m=\nu$.
We denote $m_*=m(\nu_m=\nu)$, $Z_*=Z(m_*)$,
$\tep_{B*}=\tepB(m_*)$, use $Z/Z_*=(\nu_m/\nu)^{-4/17}$, and get
\be
  L_\nu^\pm(\nu,\tR)=w(\tR)\,\frac{k}{2}\,\tep_{B*}^{1/2}\,
      \frac{Z_*}{\mu_e}\,m_*\,\left\{\frac{4}{17p-25}+\frac{6}{29}\left[
     1-\left(\frac{\nu}{\tnu_1}\right)^{29/51}\right]\right\}
  \qquad \nu\leq\tnu_1.
\label{eq:integ2}
\ee
We used again the slow variation of $m$ and $\tepB$ with $\nu_m$ and
replaced $m$ by $m_*$ and $\tepB$ by $\tep_{B*}$.

\subsection*{$\tR\geq\Racc$}

At radii $\tR>\Racc$, the blast wave material has shells with mass coordinate
$0<m<\macc$ and $\macc<m<\tm$. The $L^\pm_\nu$ integral is taken over 
$0<m<m_1=\min\{\tm,\mload\}$.

For $\nu>\tnu_1$ the integral is given by
\beq
 L_\nu^\pm(\nu,\tR) = w(\tR)\int_0^{\tnu_1} \tepB^{1/2}\frac{Z}{\mu_e}
      \left(\frac{\nu_m}{\nu}\right)^{(p-1)/2}\,\frac{k}{2}\,
     \left\{ \begin{array}{ll}
               (\xiacc/10\xi) & m>\macc\\
               2/17 & m<\macc \\
             \end{array}
     \right\} m\,\frac{\dd\nu_m}{\nu_m}. \\
\eeq
Let us denote
\be
   \tnuacc(\tR)\equiv\nu_m(\macc,\tR).
\ee
We have $Z\propto \nu_m^{-4/17}$ at $\nu_m<\tnuacc$ and 
$Z\propto\nu_m^{-1/2}$ at $\nu_m>\tnuacc$. It is convenient to calculate the 
integral as a sum of two integrals over $\nu_m<\tnuacc$ and $\nu_m>\tnuacc$.
The first integral peaks at $\nu_m=\tnuacc$ and the second at $\nu_m=\tnu_1$.
We then find
\beq
\nonumber
  L_\nu^\pm(\nu,\tR) = w(\tR)\,\frac{k}{2}
    \,\left(\frac{\nu}{\tnu_1}\right)^{(1-p)/2} \left\{ 
    \tep_{B{\rm acc}}^{1/2}\,\frac{\Zacc}{\mu_e}\,\macc\,\frac{4}{17p-25}
       \left(\frac{\tnuacc}{\tnu_1}\right)^{(p-1)/2}\right. \\
    + \left. \tep_{B1}^{1/2}\,\frac{Z_1}{\mu_e}\,m_1\,
  \frac{\xiacc}{5(p-2)\xi_1}\,
  \left[1-\left(\frac{\tnuacc}{\tnu_1}\right)^{(p-2)/2}\right]\right\}
  \qquad \nu\geq\tnu_1,
\label{eq:integ3}
\eeq
where $\Zacc\equiv Z(\macc)=74\mu_e$, $\tep_{B\rm acc}\equiv\epB(\macc,\tR)$,
and $\tep_{B1}\equiv\epB(m_1,\tR)$. 
As soon as the blast wave radius exceeds $\Racc$, we have almost 
immediately $\tnuacc\ll\tnu_1$ ($\tnu_1$ increases exponentially with $\tR$
between $\Racc$ and $\Rload$). Then the terms with $\tnuacc/\tnu_1$
can be neglected (these terms are needed only to match smoothly the 
formula with the results obtained at $\tR<\Racc$).

Next, consider intermediate $\nu$ in the range $\tnuacc<\nu<\tnu_1$.
Then the integral is calculated as a sum of three
integrals over $0<\nu_m<\tnuacc$, $\tnuacc<\nu_m<\nu$, and $\nu<\nu_m<\tnu_1$.
In the first interval we use $Z/\Zacc=(\nu_m/\tnuacc)^{-4/17}$, and 
in the other two intervals $Z/Z_*=(\nu_m/\nu)^{-1/2}$, where asterisk
refers to the point $\nu_m=\nu$. We thus get,
\beq
\nonumber
  L_\nu^\pm(\nu,\tR) & = & 
  w(\tR)\,\frac{k}{2}\left\{\frac{\Zacc}{\mu_e}\,\tep_{B\rm acc}^{1/2}\,
    \macc\,\frac{4}{17p-25}\left(\frac{\tnuacc}{\nu}\right)^{(p-1)/2} 
\right.\\ & & \hspace*{1.5cm} + \left.
\nonumber
   \frac{Z_*}{\mu_e}\,\tep_{B*}^{1/2}\,m_*\frac{\xiacc}{\xi_*}\,
\frac{1}{5(p-2)}\left[1-\left(\frac{\tnuacc}{\nu}
                        \right)^{(p-2)/2}\right]\right.\\
 && \hspace*{1.5cm} 
  + \left. \frac{Z_*}{\mu_e}\tep_{B*}^{1/2}\,m_*
   \frac{\xiacc}{\xi_*}\,
   \frac{3}{25}\left[1-\left(\frac{\nu}{\tnu_1}\right)^{5/6}\right]\right\}
   \qquad \tnuacc\leq\nu\leq\tnu_1.
\label{eq:integ4}
\eeq

Finally, at low $\nu<\tnuacc$, the integral is calculated as a sum of three
integrals over $0<\nu_m<\nu$, $\nu<\nu_m<\tnuacc$, and $\tnuacc<\nu_m<\tnu_1$.
In the first two intervals we use $Z/Z_*=(\nu_m/\nu)^{-4/17}$, and
in the last interval $Z/\Zacc=(\nu_m/\tnuacc)^{-1/2}$, 
\beq
\nonumber
  L_\nu^\pm(\nu,\tR) 
  &=& w(\tR)\,\frac{k}{2}\left\{\frac{Z_*}{\mu_e}\,\tep_{B*}^{1/2}\,
    m_*\,\frac{4}{17p-25} + \frac{Z_*}{\mu_e}\,\tep_{B*}^{1/2}\,m_*\,
\frac{6}{29}\left[1-\left(\frac{\nu}{\tnuacc}\right)^{29/51}\right]\right.\\
 && \hspace*{1.1cm} + \left. \frac{\Zacc}{\mu_e}\,\tep_{B\rm acc}^{1/2}\,
   \macc\,\frac{3}{25}\left(\frac{\nu}{\tnuacc}\right)^{1/3}
   \left[1-\left(\frac{\tnuacc}{\tnu_1}\right)^{5/6}\right]\right\}
   \quad \;\; \nu\leq\tnuacc.
\label{eq:integ5}
\eeq

\subsection*{$Z_*$ and $m_*$}

The above formulae for $L_\nu^\pm$ should be made explicit by expressing 
$m_*$, $\xi_*$, and $Z_*$ in terms of $\nu,\tR$. 

Equation~(\ref{eq:Z_nu}) gives $Z(\nu_m)$, and we get $Z=Z_*$ at $\nu_m=\nu$,
\be
\label{eq:Zstar__}
  \frac{Z_*}{\mu_e}=\left\{ \begin{array}{ll}
  \left[\nu(1+\beta_*)^{3/2}n_{0*}^{1/2}\,q^{-1}\tep_{B*}^{-1/2}\Gamma_*^{-1}
  \right]^{-1/2}               & m_*>\macc,\\
  \left[\nu(1+\beta_*)^{3/2}n_{0*}^{1/2}\,q^{-1}\tep_{B*}^{-1/2}\Gamma_*^{-1}
  \right]^{-4/17} 74^{9/17}    & m_*<\macc,\\
                   \end{array}
           \right.
\ee
All quantities with asterisk are taken at $R=R_*$ ($\nu_m=\nu$).
Since $R_*$ is very close to $\Racc$ (see eq.~\ref{eq:Rstar_} below),
it is sufficient to use the first approximation $R_*=\Racc$ in the equation 
for $Z_*$. Then we get,
\beq
\label{eq:Zstar_}
   \frac{Z_*}{\mu_e}=\left\{ \begin{array}{ll}
                    \Phi        & 1<\Phi<74, \\
  74\,(\Phi/74)^{8/17} & \Phi>74, \\
                   \end{array}
           \right.
\eeq
where
\be
\label{eq:Phi_}
 \Phi(\nu,\tR)=\left[\frac{\nu\,n_{0\rm acc}^{1/2}}
               {q\tep_{B\rm acc}^{1/2}\Gamma_{\rm acc}}\right]^{-1/2},
\ee
and the subscript ``acc'' marks that the quantity is taken at $m=\macc$.

Note that 
\be
  \Phi=74\left(\frac{\nu}{\tnuacc}\right)^{-1/2}
\ee
at $\tR\geq\Racc$.

The relation between $\xi$ and $Z$ is given by equation~(\ref{eq:Z}), 
and we find $\xi_*$,
\beq
\label{eq:xistar}
  \frac{\xi_*}{\xiacc}=\left\{ \begin{array}{ll}
                (1/5)\ln(\Phi+\sqrt{\Phi^2-1})   & 1<\Phi<74, \\
  (\Phi/74)^{4/17} & 74<\Phi<8\times 10^3, \\
                   \end{array}
           \right.
\eeq
(here $\Phi=8\times 10^3$ corresponds to $\xi_*=3\xiacc$).
Using the relation between $R$ and $\xi$ (eq.~\ref{eq:xi}), we get
\beq
\label{eq:Rstar_}
  \frac{R_*}{\Racc}=\left\{ \begin{array}{ll}
                [(1/5)\ln(2\Phi)]^{-1/2}  & 1<\Phi<74, \\
  (\Phi/74)^{-2/17} & 74<\Phi<8\times 10^3, \\
                   \end{array}
           \right.
\eeq
where we took $\Phi+\sqrt{\Phi^2-1}\approx 2\Phi$. The corresponding $m_*$ is 
\be
\label{eq:mstar_} 
  \frac{m_*}{\macc}=\left(\frac{R_*}{\Racc}\right)^k
                   =\left(\frac{\xi_*}{\xiacc}\right)^{-k/2}.
\ee

\subsection*{Final result}

The terms with $\tep_{B\rm acc}^{1/2}$ are non-negligible when $\nu\sim\nuacc$
and $m_*\approx\macc$, therefore one can replace $\tep_{B\rm acc}^{1/2}$
by $\tep_{B*}^{1/2}$ and simplify the derived expressions. 
The final result is as follows,
\be
   L_\nu^\pm(\nu,\tR)=30\,
           \left[\frac{\tn_0}{\mu_e\tg(1+\tb)}\right]^{1/2}
           \left\{ \begin{array}{ll}
  Q_*\,m_*\,Z_*\,\tep_{B*}^{1/2} & \nu\leq\tnu_1, \\
  Q_1\,m_1\,Z_1\,\epsilon_B^{1/2}\,(\nu/\tnu_1)^{(1-p)/2} 
                                            & \nu\geq\tnu_1, \\
                   \end{array}
           \right.
\ee
\beq
   Q_*(\nu,\tR) & = & \frac{k}{2}\left\{\frac{4}{17p-25}
 +\frac{6}{29}\left[1-\left(\frac{\nu}{\tnu_1}\right)^{29/51}\right]\right\}
  \qquad \tR\leq\Racc, \\
\nonumber
   Q_*(\nu,\tR)&=&\frac{k}{2}\left\{\frac{(3p-1)}{25(p-2)}\frac{\xiacc}{\xi_*}
  +\left(\frac{\tnuacc}{\nu}\right)^{(p-1)/2}
   \left[\frac{4}{17p-25}\frac{\macc}{m_*}
   -\frac{1}{5(p-2)}\frac{\xiacc}{\xi_*}\right] \right. \\
   && \left. -\left(\frac{\nu}{\tnu_1}\right)^{5/6}
   \frac{3}{25}\frac{\xiacc}{\xi_*}\right\} \hspace*{3.7cm} \tR\geq\Racc,
   \;\; \tnuacc\leq\nu\leq\tnu_1, \\
\nonumber
  Q_*(\nu,\tR) & = &\frac{k}{2}\left\{\frac{4}{17p-25}+\frac{6}{29}
  +\left(\frac{\nu}{\tnuacc}\right)^{29/51}
  \left(\frac{3}{25}\frac{\macc}{m_*}-\frac{6}{29}\right) \right. \\
  && \left. -\frac{3}{25}\frac{\macc}{m_*}
     \left(\frac{\nu}{\tnuacc}\right)^{29/51}
  \left(\frac{\tnuacc}{\tnu_1}\right)^{5/6}\right\}
  \qquad\quad \tR\geq\Racc, \;\; \nu\leq\tnuacc\leq\tnu_1, \\
 Q_1(\nu,\tR) & = & \frac{k}{2}\,\frac{4}{17p-25},\hspace*{5cm}\tR\leq\Racc,\\
\nonumber
  Q_1(\nu,\tR) & = & \frac{k}{2}\left\{\frac{1}{2(p-1)}\frac{\xiacc}{\xi_1}
   +\left(\frac{\tnuacc}{\tnu_1}\right)^{p/2-1}
    \left[\frac{4}{17p-25}\frac{\macc}{m_1} \right.\right. \\
   & & \left.\left. \hspace*{5.5cm}-\frac{1}{2(p-1)}\frac{\xiacc}{\xi_1}\right] 
       \right\} \quad \tR\geq\Racc.
\eeq
The ratios $\xi_*/\xiacc$ and $m_*/\macc$ appearing in the expressions
for $Q_*$ at $\tR>\Racc$ are given by
\beq
\label{eq:mstar}
  \left(\frac{m_*}{\macc}\right)^{-s/k}=\frac{\xi_*}{\xiacc}
   =\left\{ \begin{array}{ll}
                1-0.1\ln(\nu/\tnuacc)   & \nu\geq\tnuacc, \\
  (\nu/\tnuacc)^{-2/17} & \nu\leq\tnuacc. \\
                   \end{array}
           \right. 
\eeq

In this paper, we are interested in $\tR>\Racc$ and $\nu<\tnu_1$.
Then 
\be
  L_\nu^\pm=30\left(\tep_{B*}\tn_0\right)^{1/2}Q_*m_*Z_*,
\ee
and the numerical factor $f_\nu$ defined in equation~(\ref{eq:Lpm}) is
given by
\be
   f_\nu=Q_*\frac{m_*Z_*}{\macc\Zacc}
   \left(\frac{\tep_{B*}}{\tep_{B\rm acc}}\right)^{1/2}.
\ee
This factor is shown in Figure~9 assuming $\tep_{B*}=\tep_{B\rm acc}$.


\end{document}